\documentclass[a4paper, 11pt]{article}

    \usepackage{graphicx}
    \usepackage{ subfig}
    \usepackage{color}
    \usepackage{bm}
    \usepackage{mathrsfs}
    \usepackage{ulem}
    \usepackage{multirow}
    \usepackage{pdflscape}
    \usepackage{rotating}
    \usepackage{lscape}
    \usepackage{pdflscape}
    \usepackage{xcolor}

    \def\b{\begin{eqnarray}}
    \def\e{\end{eqnarray}}

    \def\n{\noindent}
    \makeatletter
    \newcommand*\bigdot{\mathpalette\bigdot@{.5}}
    \newcommand*\bigdot@[2]{\mathbin{\vcenter{\hbox{\scalebox{#2}{$\m@th#1\bullet$}}}}}
    \makeatother

\begin{document}

    \begin{center}
    {\huge \textbf{Integrable Cosmological Model with \vskip.3cm van der Waals Gas and Matter Creation}}

    \vspace {10mm}
    \noindent
    {\large \bf Rossen I. Ivanov and Emil M. Prodanov} \vskip.5cm
    {\it School of Mathematical Sciences, Technological University Dublin,
    Ireland,} \vskip.1cm
    {\it E-Mails: rossen.ivanov@dit.ie, emil.prodanov@dit.ie} \\
    \vskip1cm
    \end{center}

    \vskip4cm
    \begin{abstract}
    \n
    A cosmological model with van der Waals gas and dust has been studied in the context of a three-component autonomous non-linear dynamical system involving the time evolution of the particle number density, the Hubble parameter and the temperature. Due to the presence of a symmetry of the model, the temperature evolution law is determined (in terms of the particle number density) and with this the dynamical system reduces to a two-component one which is fully integrable.  The globally conserved  Hamiltonian is identified and, in addition to it, some special (second) integrals, defined and conserved on a lower-dimensional manifold, are found. The parameter choices and their implication for the global dynamics in terms of cosmological relevance are comprehensively studied and the physically meaningful parameter values are identified.

    \end{abstract}

    \newpage

    \section{Introduction}

The 2018 release of the Planck cosmic microwave background anisotropy measurements \cite{planck1} reports, consistently with previous Planck data releases, that standard single-field inflationary models with Einstein gravity, based on slow-roll concave potential $V''(\varphi) < 0$, are increasingly favoured by the data. Cosmological scalar fields (with fundamental scalar field yet to be observed) are used for the modelling of inflation, together with scalar-tensor theories, perfect fluid models, dark energy fluids interacting weakly with ordinary matter, modification of gravity, etc. The slow-roll parameters needed for an accelerated expansion in the very early Universe can be achieved through all of these. \\
On the other hand, the current acceleration of the expansion of the Universe \cite{rp} is the most mysterious aspect of cosmology today. The six-parameter base $\Lambda$CDM model, which incorporates a cosmological constant $\Lambda$, modelling dark energy, and cold dark matter is the current concordance cosmological model. It fits the observational data quite well \cite{planck2} and gives good explanation for the existence and structure of the cosmic microwave background, the large-scale structure galaxy distribution, and the abundance of elements. Perhaps the most controversial tension between the Planck $\Lambda$CDM model and astrophysical data is the discrepancy with the direct measurements of the Hubble constant --- the Planck base $\Lambda$CDM results agree well with the Baryon Acoustic Oscillations and Supernov\ae $\,$ measurements, and also with some galaxy lensing observations, but is in slight tension with the Dark Energy Survey and in significant tension with local measurements of the Hubble constant --- see \cite{planck2} and the references therein. \\
The cosmological principle, namely, that on a very large scale, the distribution of matter in the Universe is homogeneous and isotropic, leads to perfect fluids being the most common choice for the cosmological models. The equation of state of a perfect fluid connects the pressure $p$ to its energy density $\rho$ via a relationship of the type $p = \omega \rho$, where $\omega$ is independent of time. Capozziello {\it et al.} \cite{cap}
proposed the consideration of a cosmological model with the more general two-phase van der Waals fluid since a simple perfect fluid model does not describe phase transitions between successive thermodynamic states of cosmic fluids. This model also accounts for the acceleration of the expansion and is based on a binary mixture of baryons (modelled as dust) and dark matter with a van der Waals equation of state. It also allows an early de Sitter expansion, followed by a matter-dominated epoch. Consequently, Kremer \cite{kremer} proposed a binary model with van der Waals fluid and with dark energy density, the latter modelled either as quintessence or as Chaplygin gas.  Van der Waals fluid has also been used to describe the inflation of the early Universe \cite{vdwi}. \\
The Friedmann equation $\ddot{a}/a = -(4 \pi G/3)(\rho + 3 p)$ shows that $\omega < -1/3$ is required for cosmic acceleration. Separately, the energy conservation equation, $\dot{\rho}  = - 3H( \rho + p)$, shows that $\rho + p $ must not be negative so that the energy density of an expanding Universe ($H > 0$) decreases with time (realistic cosmology) which leads to the requirement $\omega  \ge  - 1$. Dark energy is defined as any physical field for which $ - 1 \le \omega < -1/3$ and which satisfies the weak energy condition $\rho \ge 0$ (positive energy density to account for the necessary density to make the Universe flat) and $\rho + p \ge 0$ (realistic cosmology). The latter is also a part of the strong energy condition. However, the part $\rho + 3p \ge 0$ of the strong energy condition must be violated to account for the needed negative pressure which drives the expansion. Dark energy is sought in many forms. The $\Lambda$CDM model corresponds to $\omega  = - 1$. Phantom cosmological models violate all four energy conditions and these models have been  increasingly favoured recently. \\
 Dark energy models are not the only models that account for the current cosmic acceleration. It may be so that the acceleration is self-driven as curvatures and length scales in the observable part of our Universe are only beginning to reach values which make the  infrared modification of gravity apparent. Thirdly, it is possible that there is some as yet undiscovered property of the gravity and matter model which accounts for the acceleration. Models with particle creation mechanism are among these. The rate of change $\Gamma$ of the particle number $N$ in co-moving volume $V$ cannot be known {\'a} priori --- it is an input quantity in the phenomenological description \cite{zimdahl2} --- and there are numerous particle creation models which investigate different forms of $\Gamma$.  The only restriction on $\Gamma$ stems from the second law of thermodynamics, which necessitates $\Gamma > 0$ so that the entropy is never decreasing: $\dot{S} / S = \dot{N} / N = \Gamma > 0$. Prigogine {\it et al.} \cite{prigo} introduced an adiabatic model with particle production in which the requested conservation of the specific entropy lead to the particle ``creation" pressure $\Pi$ being linear in the particle production rate $\Gamma$, that is: $\Pi = - (\rho + p) \Gamma / (3H)$   (one should note that the total entropy is not conserved due to the enlargement of the phase space). Here $\rho$ and $p$ are the energy density and pressure of the Universe and $H$ is the Hubble parameter. Alternative cosmological models that rely on a single pressureless fluid with a constant bulk viscosity also exist --- see \cite{zimdahl2,visc, viscosity,zimdahl}. The particle creation mechanism and the fluid viscosity are considered to have equivalent geometro-thermodynamic effects \cite{equiva} and this is manifested with the associated additional pressure term $\Pi$, called ``viscous pressure", when associated with dissipative phenomena like bulk viscosity, or ``creation pressure'' when particle number is not conserved --- see \cite{prigo, engoliam}. \\
To account for the inflation of the Universe, van der Waals fluid has also been used  \cite{brevik}, as part of a binary mixture with (pressureless) matter, in the context of existing bulk viscosity. The equation of state for the fluid considered in \cite{brevik} is $p = \omega(\rho, t) \rho + f(\rho) - 3 H \zeta(H, t)$, where $\omega(\rho, t)$ describes a time-dependent van der Waals fluid, $f(\rho)$ is an arbitrary function, and $\zeta(H, t)$ is the bulk viscosity which depends on the Hubble parameter $H$ and time. One can attribute this accumulative pressure as the effective pressure of a van der Waals fluid which itself is a carrier of viscosity. The source of the viscosity term studied in \cite{brevik} --- e.g. particle creation versus dissipative phenomena --- is not stated. In light of this, one should point out that real gases are legitimate perfect fluids, satisfying Euler equations, for as long as dissipative forces are not included, and one has $T_{\mu \nu} = (\rho + p) \, u_\mu \, u_\nu - p \, g_{\mu \nu}$. There should be no shear, stresses or heat conduction. Otherwise, a dissipative (or viscous) fluid (satisfying the Navier--Stokes equation) for which the symmetric viscosity stress tensor $\sigma_{\mu \nu} = \lambda \pi_{\mu \nu} \nabla_\rho u^\rho + v (\nabla_\mu u_\nu + \nabla_\nu u_\mu)$ is also present in $T_{\mu \nu}$, thus linearly perturbing the perfect fluid \cite{y} (here the constants $\lambda$ and $v$ are the so-called bulk viscosity and shear viscosity, respectively, and the projection tensor $\pi$ is given by $\pi_{\mu \nu} = g_{\mu \nu} + u_\mu u_\nu$).
Viscous terms however do not enter the continuity equation on the same footing as the ``creation" pressure does --- see, for example, \cite{maa}, formula (2.3). To overcome this difficulty, dissipative terms should be multiples of $H$ --- as it is done in \cite{brevik}. \\
In the present work, a two-component mixture of a real gas with van der Waals equation of state and a pressure-less dust are considered with $\rho$ and $p$ denoting the cumulative energy density and pressure. The energy density of the dust, $\rho_d$, will be allowed to take positive values (for example, one could think of the dust component in this case as of ordinary baryonic matter), to be zero (absence of dust component), or to take negative values. Dust with negative energy density is not a new feature --- see \cite{d1, d2, d3, d4, d5, d6} and the references therein. One should also mention the recently proposed model of negative masses and matter creation within a modified $\Lambda$CDM framework \cite{farnes}. \\
$\!\!$Methods from dynamical system analysis, see for example  \cite{vilasi}, \cite{ross}, are commonly used for the study of various cosmological models. With tools from \cite{nie}, this paper analyses a simple particle production model the set-up for which has been considered by many authors --- see, for example, \cite{cap}. The ``creation" pressure $\Pi$ depends only on the energy density $\rho$ and the pressure $p$, namely $\Pi = - \beta (\rho + p)$, where $\beta$ is a positive constant, that is, $\Gamma = 3 \beta H$ --- see, for example, \cite{zimdahl2, freaza}.  Clearly, this model works for the regime of expansion only, even though the regime of negative $H$ is dynamically allowed. The dynamics of the model is studied with the help of a three-component dynamical system with the particle number density $n$, the Hubble parameter $H$, and the temperature $T$ taken as dynamical variables. Due to a symmetry in the model (a first integral of the system), the temperature evolution law can be immediately determined as a function of the particle number density $n$ and, in result, the dynamical system can be easily reduced to a two-component one in terms of $n$ and $H$. Another global first integral exists (together with three second integrals) and due to it, the van der Waals dynamical system turns out to be fully integrable and having Hamiltonian structure.

    \section{The Set-up}
    The setting for the analysis is a flat Friedmann-Robertson-Walker-Lema\^itre (FRWL) cosmology with metric:
    \b
    ds^2 = g_{\mu \nu} dx^\mu dx^\nu =
    c^2 dt^2 - a^2(t) [dr^2 + r^2 (d \theta^2 + \sin^2 \theta \, d \phi^2)],
    \e
    where $a(t)$ is the scale factor of the Universe. \\
    The Universe is modelled classically as a two-component mixture. The first component is a real gas with van der Waals equation of state which can be written as a virial expansion of the pressure $p$ over the number density $n = N / V$:
    \b
    \label{eos}
    p = n T [1 + n F(T) + \ldots ].
    \e
    Here $F(T)$ denotes two-particle interaction terms (all higher-order terms, describing interactions of three or more particles, are ignored) and has the form $F(T) = A - B/T$, where $A$ and $B$ are positive constants\footnote{For illustrative purposes, the numerical example presented in this paper is for van der Waals gas with parameters $A = 1/100$ and $B = 10$.}. \\
    The second component of the Universe is taken to be dust with energy density $\rho_d$ and pressure $p_d = 0$.  \\
    In Planck units ($8 \pi G = 1, c = 1, k_B = 1$), the energy-momentum tensor $T_{\mu \nu}$, representing the two fractions of the Universe, collectively modelled with a perfect fluid, is given by:
    \b
    \label{te}
    T_{\mu \nu} = (\tilde{\rho} + \tilde{p} + \Pi) \, u_\mu \, u_\nu - (\tilde{p} + \Pi) \, g_{\mu \nu} \, .
    \e
    Here $\tilde{\rho} = \rho_d + \rho$ and $ \tilde{p} = p$ are, respectively,  the cumulative density and pressure for both fractions and $u^\mu = dx^\mu / d \tau$ (with $\tau$ being the proper time) is the flow vector satisfying $g_{\mu \nu} u^\mu u^\nu = 1$. \\
    The Friedmann equations are:
    \b
    \label{h1}
    \frac{\ddot{a}}{a} & = & - \frac{1}{6} [ \rho_d + \rho + 3 (p + \Pi) ], \\
    \label{h2}
    H^2 & = & \frac{1}{3} (\rho_d + \rho),
    \e
    where $H(t) = \dot{a}(t) / a(t)$ is the Hubble parameter. $H(t)$ will be one of the three dynamical variables of the presented model [the other
    two will be the number density $n(t)$ and the temperature $T(t)$]. As $\ddot{a} / a = \dot{H} + H^2$, combining the Friedmann equations allows to express $\dot{H}$ as  follows:

    \b
    \label{dynami}
    \dot{H} =  - \frac{3}{2} H^2 - \frac{1}{2} (p + \Pi).
    \e
    The processes of particle creation leads to non-conservation of the number of particles in the perfect fluid. This is manifested by the continuity equation: $N^\mu_{\phantom{\mu }; \mu} = n \Gamma$, where $N^\mu = n u^\mu$ is the particle flow vector and $\Gamma$ is the particle production rate.  \\
    The particle conservation equation can be written as
    \b
    \label{nn}
    \dot{n} = - 3 n H + \Psi,
    \e
    where $\Psi = n \Gamma$. This equation will be further used as one of the evolution equations in a dynamical system of three simultaneous autonomous differential equations [in terms of the number density $n(t)$, the Hubble parameter $H(t)$, and the temperature $T(t)$]. \\
    Many forms of the term $\Psi$ have been considered in the literature. This paper studies the dynamics of an expanding Universe with particle creation term in the form \cite{freaza}:
    \b
    \label{Psi}
    \Psi = 3 \beta n H,
    \e
    where $\beta$ is a positive constant which will be treated as parameter of the model. \\
    The energy conservation equation for the real gas is:
    \b
    \label{h3}
    \dot{\rho} + 3H(\rho + p + \Pi) = 0
    \e
    and that of the dust is:
    \b
    \label{ce-d}
    \dot{\rho_d} + 3H \rho_d  = 0
    \e
    --- by taking these two equations separate from each other, a choice is made that there would be no exchange between the two fractions. \\
    To find an expression \cite{prigo} for the ``creation pressure" $\Pi$, consider the Gibbs equation:
    \b
    \label{gibbs}
    T ds = d \Bigl({\rho \over n}\Bigr) + p \, d \Bigl({1\over n}\Bigr) =
    -\left({\rho + p \over n^2}\right) \,\, dn + {1 \over n} \,\, d\rho.
    \e
    Here $s$ is the specific entropy (entropy per particle, $s = S/N$, where $S$ is the total entropy and $N$ --- the total number of particles). In the above, $T$ is the temperature of the Universe. \\
    With the help of the particle conservation equation (\ref{nn}) and the continuity equation (\ref{h3}), the Gibbs equation becomes:
    \b
    n T \dot{s} = - (\rho + p) \frac{\dot{n}}{n} + \dot{\rho} = - 3 H \Pi - \Gamma (\rho + p).
    \e
   If the specific entropy is conserved, one immediately finds \cite{prigo}:
    \b
    \label{Pi}
    \Pi = -\frac{\Gamma (\rho + p)}{3H} = - \frac{\rho + p}{n} \frac{\Psi}{3H}.
    \e
    Note that the total entropy $S$ is not conserved due to the enlargement of the phase space resulting from the particle production \cite{prigo}. \\
    The energy conservation equation thus becomes:
    \b
    \label{h33}
    \dot{\rho}(n, T) = - 3H(\rho + p) \left(1 - \frac{\Psi}{3nH}\right).
    \e
    Substituting (\ref{Psi}) into (\ref{dynami}) yields:
    \b
    \label{hashdot}
    \dot{H} =  - \frac{3}{2} H^2 - \frac{1}{2}
    \left[
    (1 - \beta) p(n,T) - \beta \rho(n, T) \right].
    \e
    This is the dynamical evolution equation for the Hubble parameter and the second equation of the dynamical system of three simultaneous autonomous  differential equations. \\
    The particle conservation equation (\ref{nn}) can be re-written as $a^3 \dot{n} + 3 a^2 \dot{a} n = a^3 \Psi.$ Thus: $(d/dt) (a^3 n)  = dN/dt = a^3 \Psi$ and
    \b
    n(t) = \frac{1}{a^3(t)} \int\limits_{t_0}^t a^3(t') \Psi(t') dt'.
    \e
    On the other hand, differentiating $N = n a^3$ with respect to time, using $\dot{a} = a H$ and (\ref{nn}), yields:
    \b
    \label{dotn}
    \dot{N} = 3 \beta N H.
    \e
    Separately, differentiating the specific entropy $s = S/N$ with respect to time and using the fact that it is conserved ($\dot{s} = 0$), one can immediately find:
    \b
    \label{entr}
    \frac{\dot{S}}{S} = \frac{\dot{N}}{N} = 3 \beta H
    \e
    and thus, for a model with increasing entropy, one can only consider regime of cosmic expansion ($H = \dot{a} / a > 0$), and not for contraction as $H < 0$, despite being dynamically allowed, leads to decreasing entropy and violation of the second law of thermodynamics. \\
    Noting that the specific entropy is a full differential, the Gibbs equation (\ref{gibbs}) yields the following integrability condition (in chosen thermodynamical variables $\rho$ and $n$):
    \b
    \biggl[ \frac{\partial}{\partial n}
    \biggl(\frac{\partial s}{\partial \rho}\biggr)_n\biggr]_\rho =
    \biggl[ \frac{\partial}{\partial \rho}
    \biggl(\frac{\partial s}{\partial n}\biggr)_\rho\biggr]_n
    \qquad
    \mbox{or} \qquad \biggl[\frac{\partial}{\partial n} \biggl(\frac{1}{Tn}\biggr)\biggr]_\rho =
    \biggl[\frac{\partial}{\partial \rho} \biggl(-\frac{\rho + p}{Tn^2}\biggr)\biggr]_n.
    \e
    This can be written as
    \b
    \label{integrability}
    n \biggl(\frac{\partial T}{\partial n}\biggr)_\rho +
    (\rho + p)\biggl(\frac{\partial T}{\partial \rho}\biggr)_n
    = T
    \biggl(\frac{\partial p}{\partial \rho}\biggr)_n.
    \e
    For any simple thermodynamical system, one has the following relationship \linebreak $(\partial Z / \partial \zeta)_T \,\, (\partial \zeta / \partial T)_Z \,\,(\partial T / \partial Z)_\zeta = -1$, where $Z$ is the acting generalized force, associated with the external parameter $\zeta$, i.e. $Z = Z(\zeta, T)$ (this is the thermic equation of state and it is warranted by the second initial proposition of thermodynamics). The integrability condition can therefore be written as the following thermodynamic identity:
    \b
    \label{tdi}
    \rho + p = T \biggl( \frac{\partial p}{\partial T}\biggr)_n + n \biggl( \frac{\partial \rho}{\partial n}\biggr)_T.
    \e
    In thermodynamical variables $n$ and $T$, the dynamics of the energy density is given by:
    \b
    \dot{\rho}(n, T) = \biggl(\frac{\partial \rho}{\partial n}\biggr)_T \dot{n}
    + \biggl(\frac{\partial \rho}{\partial T}\biggr)_n \,\, \dot{T}.
    \e
    Substituting the number conservation equation (\ref{nn}) and the energy conservation equation (\ref{h33}) gives:
    \b
    - 3H(\rho + p) \left(1 - \frac{\Psi}{3nH}\right) = (\Psi - 3nH) \, \biggl(\frac{\partial \rho}{\partial n}\biggr)_T
    + \biggl(\frac{\partial \rho}{\partial T}\biggr)_n \,\, \dot{T}.
    \e
    Using the thermodynamic identity (\ref{tdi}) to replace the term $\rho + p$ on the left-hand side in the above, one immediately finds the following temperature evolution law:
    \b
    \label{T}
    \dot{T} = \left(\frac{\Psi}{n} - 3H \right) T
    \biggl( \frac{\partial p}{\partial \rho} \biggr)_n = \left( \frac{\Psi}{n} - 3H \right) T \frac{\left( \frac{\partial p}{\partial T} \right)_n}{\Bigl( \frac{\partial \rho}{\partial T} \Bigr)_n} .
    \e
    This is the third dynamical equation. \\
    In the absence of particle creation (i.e. when $\Psi = 0$), the above reduces to the well known form given in \cite{nie,maa, lima}. \\
    Using the equation of state (\ref{eos}) for the van der Waals gas, namely:
    \b
    \label{pe}
    p(n,T) = n T (1 + An) - B n^2,
    \e
    one finds $(\partial p / \partial T)_n = n(1 + An).$
    Substituting this, together with the equation of state, into the integrability condition (\ref{integrability}) gives the following differential equation:
    \b
    \biggr[ \frac{\partial}{\partial n} \biggl( \frac{\rho}{n} \biggr) \biggr]_T = - B.
    \e
    This integrates directly into:
    \b
    \rho = n [ \phi(T) - B n],
    \e
    where $\phi(T)$ can be determined as follows. Consider an ideal gas limit (by setting the coefficient $F(T) = A - B/T$ of the second term of the virial expansion to zero). In the case of monoatomic gas with three translational degrees of freedom, the average kinetic energy of the particles is $(3/2) T$. Also, $n = (\mathcal{N} m) / (V m) = (M / V) (1/m) = \rho / m$, where $M$ is the mass of the system and $m$ is the relativistic mass of a representative particle\footnote{For the numerical example in this paper, the value chosen for $m_0$ is $100$.}: $m = m_0 + (1/2) m_0 u^2 + O(u^4).$ Here $m_0$ is the rest mass and $u$ --- the speed of the particle. One can write the mass density of the ideal gas approximately as $\rho = n [m_0 + (3/2) T]$. Thus, $\phi(T) = m_0 + (3/2) T$.
    One immediately finds the relationship between the number density $n$, the mass density $\rho$ and the temperature $T$ of the van der Waals gas:
    \b
    \label{rvg}
    \rho(n, T) = n (m_0 + \frac{3}{2} T) - B n^2.
    \e
    Thus, $(\partial \rho / \partial T)_n = (3/2) n$ and the temperature law (\ref{T}) for the van der Waals gas becomes $\dot{T} = - 2 [ H - \Psi/(3n) ] T (1 + A n)$. \\
    Finally, the resulting dynamical system for the case of a van der Waals gas is:
    \b
    \label{1}
    \!\!\dot{n} \!\!\! & = & \!\!\!\! 3 (\beta - 1) n H, \\
    \label{2}
    \!\!\dot{H} \!\!\! & = & \!\!\!\! - \frac{3}{2} H^2 + \frac{1}{2} \left[ (\beta - 1) p(n, T)  + \beta  H \rho(n, T) \right],  \\
    \label{3}
    \!\!\dot{T} \!\!\! & = & \!\!\!\!  2 (\beta - 1) (1 + A n) H T,
    \e
    where $p(n, T) =  n T ( 1 + A n ) - B n^2$ and $\rho(n, T) =  n [m_0 + (3/2) T] - B n^2$. \\
    There is a symmetry in the model: if one divides (\ref{3}) by (\ref{1}), an expression independent of $H$ stems:
    \b
    \label{asb}
    \frac{dT}{dn} = \frac{2T(1+An)}{3n} > 0 \,
    \mbox{ as $n > 0$.}
    \e
    The solution is given by the monotone continuous function:
    \b
    \label{t}
    T(n) = \tau \, n^{\frac{2}{3}} \, e^{\frac{2 A n}{3}},
    \e
    where $\tau$ is a positive constant. It represents a temperature scale which will be treated as a parameter of the model (together with the other parameter $\beta$). \\
    Equation (\ref{asb}) and its solution are the same as the ones encountered in the case of absence of matter creation \cite{nie}. \\
    There is a {\it global} first integral given by:
    \b
    \label{i1}
    I_{1}(n,T) = T \, n^{-\frac{2}{3}} \, e^{-\frac{2 A n}{3}} = \tau \mbox{ = const } > 0.
    \e
    Using (\ref{t}), the temperature can be excluded from the system to give:
    \b
    \label{2d-1}
    \dot{n} & \equiv f_1(n, H) & = 3 (\beta - 1) n H,  \\
    \label{2d-2}
    \dot{H}  & \equiv f_2(n, H) & =
    \, - \frac{3}{2} \, H^2 - \frac{1}{2} \, \tau \, n^{\frac{5}{3}} \, e^{\frac{2An}{3}} \left[ (1 - \beta) \left(\frac{5}{2} + An \right) - \frac{3}{2} \right] \nonumber \\
    && \hskip3.2cm + \,\, \frac{1}{2} \, \beta \, (m_0 - 2Bn)\, n \,\, + \,\,\frac{1}{2} \, B \, n^2
    \e
    and this resulting two-component dynamical system will become the focus of attention. \\
    Eliminating the temperature dependence of the energy density (\ref{rvg}) with the help of (\ref{t}) yields:
    \b
    \rho[n, T(n)] = n (m_0 + \frac{3}{2} \tau \, n^{\frac{2}{3}} \, e^{\frac{2 A n}{3}}) - B n^2.
    \e
    A {\it second integral} $K(\vec{x}) = 0$ of an autonomous dynamical system of the type $\dot{\vec{x}}(t) = \vec{F}[\vec{x}(t)]$ is defined as an invariant, but only on a restricted subset, given by its zero level set \cite{gor}. It is defined by $(d/dt) K(\vec{x}) = \mu(\vec{x}) K(\vec{x})$. If a trajectory starts on such invariant manifold, it remains on it throughout its evolution. This means that no trajectory can cross a hyper-surface defined by a second integral. \\
    For the three-component dynamical system, the hyper-surface, defined by $K_1 = n = 0$, is one such invariant manifold, i.e. $n = 0$ is a second integral since $(d/dt) n = [- 3 (1 - \beta) H] n$. The surface defined by $K_2 = 3H^2 - \rho = 3H^2 - n[m_0 + (3/2)T] + Bn^2 = 0$ is another second integral because $(d/dt)(3H^2 - \rho) = -3H(3H^2 - \rho).$ It is a {\it separatrix} --- see Figure 1. Similarly, the hyper-surface $K_3 = 0$, defined by $T = 0$, is another second integral and invariant manifold since $(d/dt)T = [2 (\beta - 1) (1 + A n)] H T$. \\
    As it will be necessary for the forthcoming analysis, one needs to determine at what value $\tau = \tau_0$ the separatrix $3H^2 - n[m_0 + (3/2)T] + Bn^2 = 3H^2 - n[m_0 + (3/2) \, \tau \, n^{2/3} \, e^{2An/3}] + Bn^2 = 0$ is tangent to the $n$-axis and at what point $n_0$ this happens. When $\tau = \tau_0$, the separatrix has a minimum at $n_0$ and that minimum is $0$ (see Figure 1). Thus, $(3/2) \, \tau_0 \, n_0^{2/3} \, e^{2An_0/3} = Bn_0 - m$ and $(d/dn) \left[ n [m_0 + (3/2) \, \tau \, n^{2/3} \, e^{2An/3}] - Bn^2\right]_{n=n_0, \tau = \tau_0}$ $= 0$. From these two simultaneous equations, one can immediately determine that the separatrix $K_2 = 0$ is tangent to the $n$-axis at $n_0 = [ 2 m_0 A + B   + (4 m_0^2 A^2 + 20 m_0 A B  + B^2)^{1/2} \, ] / (4AB)$ (the other root of the resulting quadratic equation is irrelevant as it is negative), provided that $\tau = \tau_0 = (2/3) (Bn_0 - m_0) n_0^{-2/3} \, e^{-2An_0/3}$. \\
    \begin{figure}[!ht]
    \centering
    \includegraphics[height=6cm,width=0.5\textwidth]{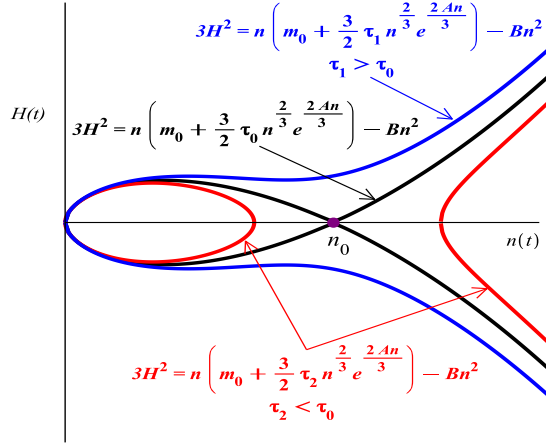}
    \caption{\footnotesize{The second integral ({\it separatrix}) $K_2 = 3H^2 - n[m_0 + (3/2) \, \tau \, n^{2/3} \, e^{2An/3}] + Bn^2 = 0$. It is an open curve when $\tau > \tau_0$ (where $\tau_0 \approx 14.78$ for a van der Waals gas with parameters $A = 0.01$ and $B = 10$ and for $m_0 = 100$), while, when $\tau < \tau_0$, it exhibits a loop at small number densities, together with an open curve at higher $n$.  When $\tau > \tau_0$, the trajectories to the right of the open curve correspond to a dust component with negative energy density $\rho_d$, while those to the left have $\rho_d > 0$.  When $\tau < \tau_0$, the trajectories to the right of the open curve and those inside the loop correspond to a dust component with negative energy density $\rho_d$ and the rest have $\rho_d > 0$. The curve with $\tau = \tau_0$ is tangent to the abscissa at $n_0 = \bigl( 2 m_0 A + B   + \sqrt{4 m_0^2 A^2 + 20 m_0 A B  + B^2} \, \bigr) / (4AB)$ (for the numerical example, $n_0 = 73.59$). The energy density $\rho[n, T(n)] = n[m_0 + (3/2) \, \tau \, n^{2/3} \, e^{2An/3}] - Bn^2$ is positive for all values of $n$ if $\tau > \tau_0$. }}
    \label{F1}
    \end{figure}
    Depending on the initial conditions (the choice of $\tau$), the trajectories for which the energy density $\rho[n, T(n)] = n[m_0 + (3/2) \, \tau \, n^{2/3} \, e^{2An/3}] - Bn^2$ is positive for all values of $n$, are those with $\tau > \tau_0$, while for values of $\tau$ below $\tau_0$, the energy density $\rho[n, T(n)]$ becomes negative over a finite region of positive values of $n$ (see Figure 1). Thus, such trajectories would become un-physical in this range for $n$ or, in fact, they could be admitted as trajectories exhibiting temporary violation of the weak energy condition --- admissible in phantom cosmology models \cite{car}. In the latter sense, the validity of the model will be extended to include large values of the number density $n$. \\
    The stability matrix $L$ for the two-component dynamical system (\ref{2d-1})--(\ref{2d-2}) is given by:
    \b
    L_{11} & = & \frac{\partial f_1}{\partial n} = 3 (\beta - 1) H, \\
    L_{12} & = & \frac{\partial f_1}{\partial H} = 3 (\beta - 1) n, \\
    L_{21} & = & \frac{\partial f_2}{\partial n} = \frac{1}{3} \, \tau \, n^{\frac{2}{3}} \, e^{\frac{2An}{3}} \,  \left[ ( \beta - 1) \left( \frac{5}{2} + An \right)^2
    + \frac{3}{2} \, \beta A n + \frac{15}{4} \right]
    \nonumber \\
    && \hskip1.2cm
    + \,\, \frac{1}{2} \, \beta m_0  + (1 - 2 \beta)Bn, \\
    L_{22} & = &  \frac{\partial f_2}{\partial H} =  - 3H.
    \e
Returning to the three-component dynamical system, one notes that integration of the dust conservation equation (\ref{ce-d}) yields $\rho_d = \rho_{d, 0} \, \exp[3 \int_{t_0}^t H(t')dt']$. From the Friedmann equation (\ref{h2}), one can express the dust density $\rho_d$ as $3H^2 - \rho$. Additionally, integration of the dynamical equation (\ref{1}) gives $\exp[3 \int_{t_0}^t H(t')dt'] = C' n^{1/(\beta - 1)}$, where $C'$ is a positive constant. Finally, using (\ref{rvg}) to eliminate $\rho$, yields another {\it global} first integral of the system:
\b
\label{i2}
I_2(n, H, T) = \Bigl[ 3 H^2 - n \Bigl( m_0 + \frac{3}{2} T \Bigr) + B n^2 \Bigr] n^{\frac{1}{\beta - 1}} = C \mbox{ = const }
\e
or
\b
\label{i2_2}
I_2(n, H) = \Bigl[ 3 H^2 - n \Bigl( m_0 + \frac{3}{2} \, \tau \, n^{\frac{2}{3}} \, e^{\frac{2An}{3}} \Bigr) + B n^2 \Bigr] n^{\frac{1}{\beta - 1}} = C \mbox{ = const }
\e
for the two-component system (\ref{2d-1})--(\ref{2d-2}). \\
Separately, since $K_1 = n = 0$ is a second integral, no trajectory can reach a point on the $H$-axis, including the origin, unless the trajectory starts on the $H$-axis itself, i.e. if the trajectory is with $n_0 = 0$. \\
The second integral
\b
K_2 = 3H^2 - \rho [n, T(n)] = 3H^2 - n[m_0 + (3/2) \, \tau \, n^{2/3} \, e^{2An/3} + Bn^2 = 0
\e
represents the trajectory with absent dust component ($\rho_d = 0$), that is, $K_2$ is equal to the first integral $I_2$ with $C = 0$. Due to the existence of a second integral, the phase space is fragmented into separate regions, each with a specific regime of $\rho_d$, thus the curve $K_2 = 0$ is called a {\it separatrix} --- see Figure 1. \\
One should note that, due to the presence of the two first integrals, the three-dimensional system can be reduced to one equation. Formally, from (\ref{i2_2}), one has:
\b
H(n) = \pm \, \frac{1}{3} \, \sqrt{n \left[ m_0 + \frac{3}{2} \, T(n)\right] - B n^2 + C n^{\frac{1}{1 - \beta}}}
\e
and then, from (\ref{1}):
\b
\int\limits_{n_0}^{n} \, \frac{d \tilde{n}}{\tilde{n} \, H(\tilde{n})} \, \, = \, \, 3 \, (\beta-1) \, (t-t_0)
\e
or
\b
\int\limits_{n_0}^{n} \, \frac{d \tilde{n}}{\tilde{n} \, \sqrt{\tilde{n} \left[ m_0 + \frac{3}{2} \, \tau \, \tilde{n}^{\frac{2}{3}} \, e^{\frac{2 A \tilde{n}}{3}}\right] - B \tilde{n}^2 + C \tilde{n}
^{\frac{1}{1 - \beta}}}} \, \, = \, \, \pm \, (\beta - 1) \, (t - t_0).
\e
The integral on the left-hand side defines some function, say $\xi$ of $n$,  which, however, also depends on the following parameters: $m_0, \tau, A,B,C$ and the initial condition $n_0$, i.e.
\b
\xi(n; m_0, \tau, A,B,C, n_0) \,\, = \,\, \pm \, (\beta - 1) \, (t - t_0).
\e
The function $\xi$ is probably impossible to find explicitly or, even if possible, given that it depends on so many parameters, one is likely to expect that its form and behaviour would strongly depend on the relationship between these parameters. And this is only half of the trouble. One does not need $t = t(n)$ but, rather, $n(t) = \xi^{-1}(t; \beta, m_0, \tau, A,B,C, n_0, t_0)$, i.e. the inverse of the function $\xi$. Not only this, one would then have to find $H(t) = H[\xi^{-1}(t; \beta, m_0, \tau, A,B,C, n_0, t_0)]$ --- a task hardly achievable even numerically. Even if the function $\xi^{-1}$ was known somehow in terms of special or, even, elementary functions, the formal solution presented above is of little or no practical relevance, since it is impossible to see or analyze its behavior. \\
Instead, the phase-space dynamical analysis of the two-component system (\ref{2d-1})--(\ref{2d-2}), as always, reveals all the essential information about the global behavior of the system. The fact that the system is Hamiltonian is a bonus which facilitates the analysis. Many different scenarios stem from the fact that one is dealing with several model parameters (a table with references to the phase portraits, provided at the end, summarizes all interrelations between the model parameters leading to different types of global behaviour). Furthermore, due to the presence of second integrals (separatrices), several types of trajectories are separated by these invariant curves. This corresponds to differences in the global behavior, depending on the initial conditions $(n_0, H_0).$ Thus, the whole complexity of the global behavior, reflecting the multitude of choices for the {\it parameters} and {\it initial data}, can be only be captured and explained through phase-space analysis. \\
As the energy density of the dust can be positive, zero, or negative, in line with this, the first integral $I_2(n, H) = C$ will be allowed to be positive, zero, or negative. \\
Due to the presence of the first integral $I_2$, the two-component van der Waals system is {\it fully integrable and has Hamiltonian structure}. To illustrate this, introduce:
    \b
    \label{46}
    u(n) & = & \frac{2}{3} n^{-\gamma}, \\
    v(n, H) & = & H n^{-\gamma},
    \e
    with $\gamma = [2(1 - \beta)]^{-1}$. The two-component system (\ref{2d-1})--(\ref{2d-2}) becomes:
    \b
    \dot{u} & = &   v, \\
    \dot{v} & = & \varphi(u)
    \e
    where
    \b
    \varphi(u) = -\frac{1}{2} n^{-\gamma} [(1-\beta)p(n) -\beta \rho(n)]
    \mbox{ \,\,\, and \,\,\, $n = n(u)$ \,\,\, from \,\,\,   (\ref{46}).}
    \e
    To re-write the above in terms of the canonical variables, consider the following. The first integral $I_2 (n, H) =$ const is the only conserved quantity for the two-component system and one would expect the Hamiltonian $\mathcal{H}(u,v)$ (conserved quantity) to be related to $I_2$. One would further guess that $\mathcal{H}(u,v) = (1/6) I_2$ in order to get a ``proper" kinetic energy term $(1/2) v^2$. It is easy to see that such guess is correct.
    \b
    \mathcal{H}(u,v) = \frac{1}{6} I_2 = \frac{1}{6}  n ^{-2\gamma} (3H^2 - \rho) = \frac{1}{2} v ^2 -\frac{3u^2}{8}\rho(u) = \frac{1}{2}v ^2 + V(u).
    \e
    Then:
    \b
    \dot{u} & = & \frac{\partial \mathcal{H}}{\partial v} = v, \\
    \dot{v} & = & -\frac{\partial \mathcal{H}}{\partial u} = \varphi(u).
    \e
    It is not difficult to check that
    \b
    \frac{\partial \mathcal{H}}{\partial u} = -\varphi(u) = \frac{1}{2} n^{-\gamma}[(1-\beta)p -\beta \rho],
    \e
    that is,
    \b
    \frac{d}{d u} V(u) = \frac{d}{d u} \left[ -\frac{3u^2}{8} \rho(u) \right] =  \frac{1}{2}n^{-\gamma}[(1-\beta)p -\beta \rho]
    \e
    or
    \b
    \frac{u}{2} \frac{d \rho}{du} = -(1-\beta)(p+\rho)=-\frac{1}{2\gamma}(p+\rho)
    \e
    Since $d \rho / du = (d\rho / dn) (dn / du)$, one gets:
    \b
    \frac{d \rho}{dn} = -\frac{2}{u}\frac{du}{dn} \frac{1}{2\gamma} (p + \rho) = -\frac{1}{\gamma} (p + \rho) \frac{d}{dn} \ln u.
    \e
    Noting that
    \b
    \frac{d}{dn} \ln u = -\gamma \frac{1}{n},
    \e
    the above yields:
    \b
    \frac{d \rho}{dn} = \frac{p + \rho}{n}.
    \e
    That this is indeed the case can be easily seen from (\ref{gibbs}) --- the specific entropy $s$ is conserved. Therefore $\mathcal{H}(u,v) = (1/6) I_2$ indeed.\\
    As the two-component system (\ref{2d-1})--(\ref{2d-2}) is Hamiltonian, the critical points are either saddles (with real eigenvalues of the stability matrix, i.e. $\lambda_{1,2} = \pm q$) or centres (with purely imaginary eigenvalues $\lambda_{1,2} = \pm i \omega$). \\
   To determine the critical points of the two-component dynamical system (\ref{2d-1})--(\ref{2d-2}), revisit the three-component system (\ref{1})--(\ref{3}) and consider first $H = H^* = 0$ in it. If, further,
    \b
    \label{qq1}
    (1- \beta) p[n^*, T^*(n^*)] - \beta \rho[n^*, T^*(n^*)] = 0,
    \e
    then the right-hand-sides of all three equations in the dynamical system vanish. Solving (\ref{qq1}) for $T^*(n^*)$ results in the following critical points:
    \b
    \label{star}
    \Biggl(n^*, \,\, H^* = 0, \,\, T^*(n^*) = \frac{(2 \beta - 1) B n^* - \beta m_0}{(\beta - 1) A n^* + \frac{5}{2} \beta - 1}\Biggr).
    \e
    Which particular values of $n^*$ (and, hence, $T^*$) the system will choose depends on the initial conditions (together with the parameters of the model) and this is manifested by the presence of the global first integrals. For the initial condition $(n_0, T_0)$ at initial time $t = t_0, \,\,$  one has $I_1(n, T) = I_1(n_0, T_0) = \tau = $ const. On the other hand, the curve $T^*(n^*)$ intersects the hyper-surface given by $I_1(n, T) = \tau$ exactly at points with coordinates $(n^\ast, T^\ast)$, satisfying (\ref{star}), namely, the equation $I_1(n^\ast, T^\ast) =  I_1(n_0, T_0) = I_1(n, T) = \tau = $ const. Together with equation (\ref{qq1}), these are the two simultaneous equations selecting the particular critical points of the type (\ref{star}) that the system will encounter for the chosen initial conditions $(n_0, T_0)$, i.e. choice of the constant $\tau$. Depending on the values of $\beta$ and $\tau$, the number of intersection points of $T^*(n^*)$ with $T(n) = \tau n^{2/3} \exp(2An/3)$, that is, the number of critical points of type (\ref{star}), could be one, two, or three (see Figure 2). \\
    The coordinates of the critical points of the type (\ref{star}) for the two-component system are alternatively given by $H^* = 0$ and $n^*$ being the solutions of:
    \b
    \tau \, n^{*^{\frac{2}{3}}} \, e^{\frac{2An^*}{3}} \left[( \beta - 1) \left(\frac{5}{2} + An^* \right) + \frac{3}{2} \right]
    - (2 \beta - 1) B n^* + \beta m_0 = 0.
    \e
    The energy density at the critical point is given by
    \b
    \rho^* \equiv \rho[n^*,  T^*(n^*)] = \frac{(\beta - 1) n^* \left[ - ABn^{*^2} + \left( m_0 A + \frac{B}{2} \right) n^* + m_0 \right]}{(\beta - 1)An^* + \frac{5}{2} \beta - 1}
    \e
    As discussed, the energy density can be temporarily negative when $\tau < \tau_0$. \\
    When $\tau < \tau_0$ and for $\beta > 1$ and also for $0 < \beta < 2/5$, one has $\rho^* > 0$ for $0 \le n^* \le N^*$, where
    \b
    N^* = \frac{m_0 A + \frac{B}{2} + \frac{1}{2} \sqrt{4 m_0^2 A^2 + 20 m_0 A B + B^2}}{2AB}
    \e
    (in the numerical example, $N^* = 73.5890$). For the case of $2/5 < \beta < 1/2$, one has $\rho^* > 0$ for $(5/2 \beta - 1)/[(1 - \beta)A] \le n^* \le N^*$, while for $1/2 < \beta < 1$, one has $\rho^* > 0$ for $n^*$ between $(5/2 \beta - 1)/[(1 - \beta)A]$ and $N^*$ (depending on $\beta$, the former can be bigger or smaller than the latter). \\
    The eigenvalues of the stability matrix at the critical points (\ref{star}) satisfy:
    \b
    \label{eig}
    \lambda^{*^2} & = & 3 (\beta - 1) n^* L_{21}^* \nonumber \\
    & = & (\beta -1) \, n^* \, \left\{ T^*(n^*) \,  \left[ ( \beta - 1) \left( \frac{5}{2} + An^* \right)^2
    + \frac{3}{2} \, \beta A n^* + \frac{15}{4} \right]
    \right.
    \nonumber \\
    && \hskip2.98cm \Biggl.
    + \,\, 3 \beta m_0  + 3(1 - 2 \beta)Bn^*  \Biggr\}.
    \e
    It is clear that the eigenvalues are either real (when $\beta -1$ and $L_{21}^*$ have the same signs) or purely imaginary (when $\beta -1$ and $L_{21}^*$ have opposite signs). This is not a surprise as Hamiltonian systems can only have centres or saddles. The different regimes of the parameters $\beta$ and $\tau$ and the resulting critical points are shown on Figures 2, 3, and 4. See also the Table for a summary of all possible cases where Figures 2, 3, and 4 are referenced in detail. \\
 \begin{figure}[!ht]
    \centering
    \subfloat[\scriptsize  Positive values for $T^*(n^*)$ exist for $n^* > \beta m_0 / \bigl( (2 \beta - 1) B \bigr)$. The function $T^*(n^*)$ increases monotonously from $\beta m_0 / (5\beta / 2 - 1)$ when $n^* = 0$ and tends to $(2 \beta - 1) B / \bigl( (\beta - 1)A \bigr)$ as $n \to \infty$. When $\tau > \tau_0$, there are no intersection points between the curves $T^*(n^*)$ and $T(n)$. In this case, the origin is the only critical point --- see Figure 2b. If $\tau < \tau_0$, then, in addition to the origin, there are two more critical points --- a centre and a saddle --- see Figure 2c.]
    {\label{F2a}\includegraphics[height=4.1cm, width=0.32\textwidth]{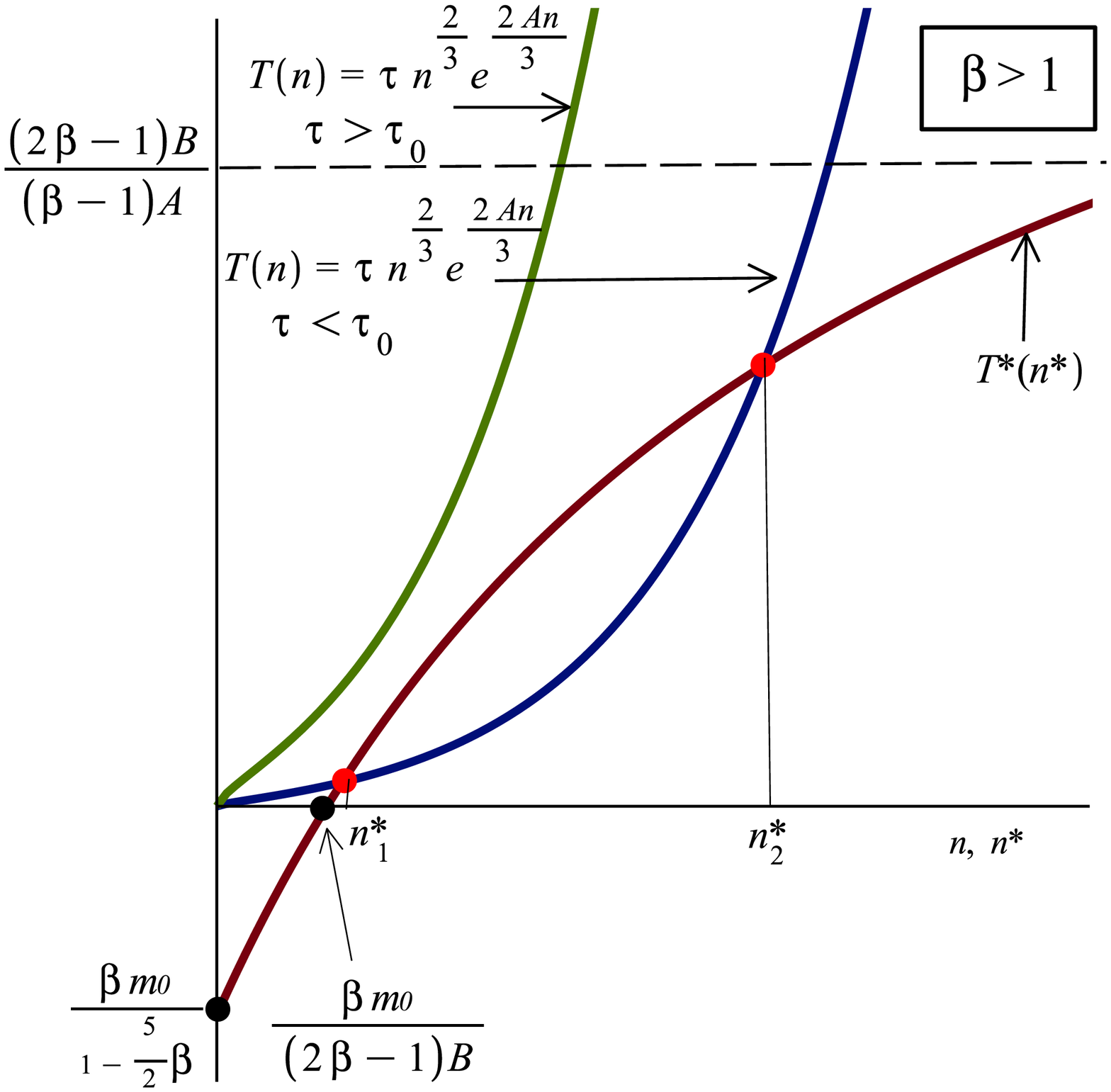}}
    \,
    \subfloat[\scriptsize As $T^*(n^*)$ and $T(n)$ do not intersect when $\tau = 15.0$, (that is, when $\tau > \tau_0$) and $\beta = 1.2$ (that is, $\beta > 1$), the origin is the only critical point. It repels all trajectories, except those on the second integral $n=0$ with $H_0 > 0$ and the separatrix itself (which is another second integral). The origin in reachable along these curves in infinite time. The physical trajectories are all those for which $H_0 > 0$. These trajectories diverge to $H \to \infty$ and $n \to \infty$. All physical trajectories become very close to the separatrix when $H$ and $n$ are very large. In this case, the leading term in $T(n)$ grows exponentially with $n$. Then $3H^2 \sim \rho \sim (3/2)nT$, also $p \sim A n^2 T > 0, \,\, \dot{H} = (\beta-1)(3H^2 +p)/2 > 0$. Thus $\ddot{a}/a= \dot{H} + H^2 >0$ and this region is characterised by inflation. ]
    {\label{F2b}\includegraphics[height=4.1cm,width=0.32\textwidth]{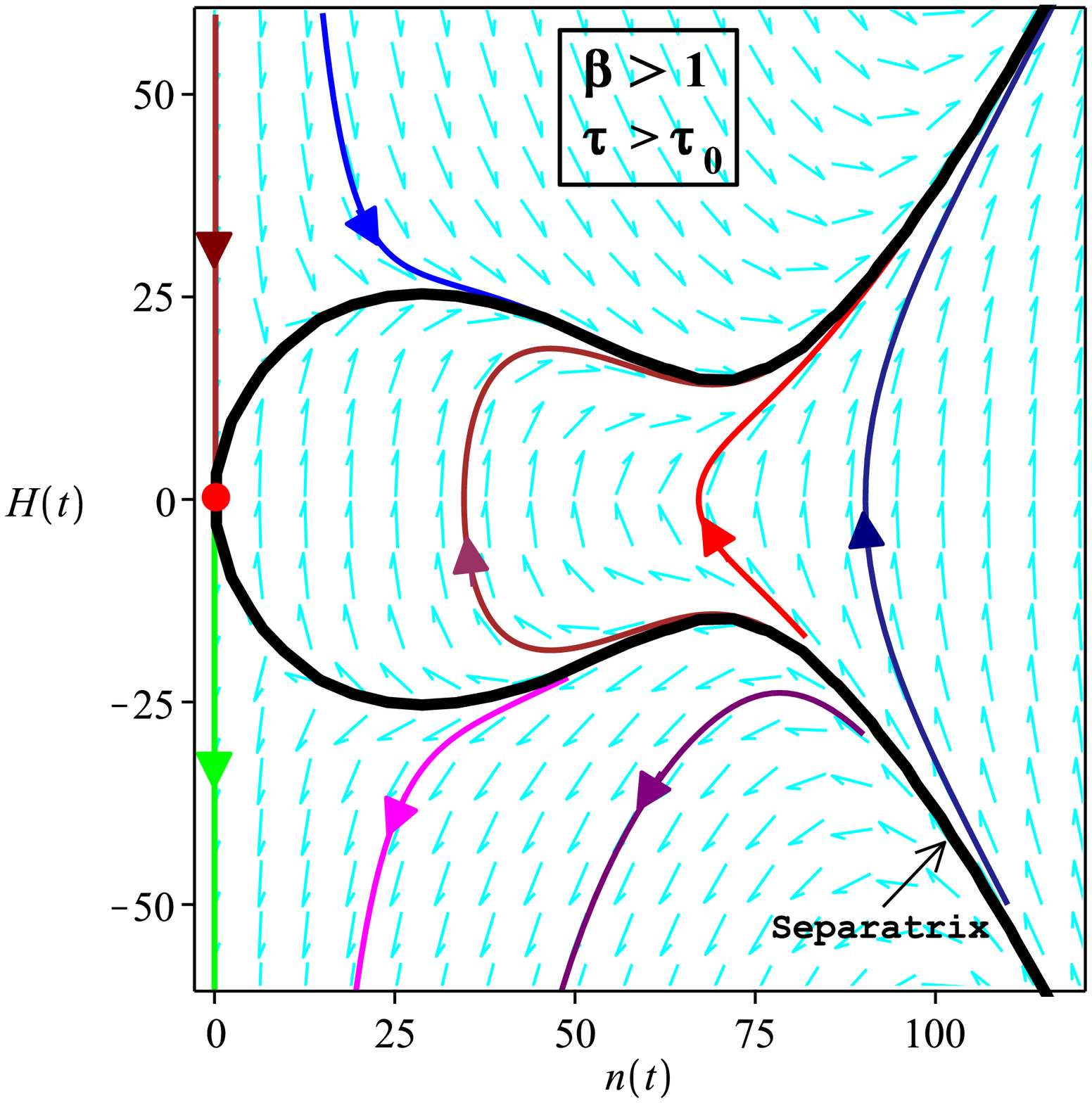}}
    \,
    \subfloat[\scriptsize When $\tau = 14.0$ (that is, $\tau < \tau_0$) and $\beta = 1.2$ (that is, $\beta > 1$), there are two intersections of  $T^*(n^*)$ and $T(n)$. This leads to the existence of three critical points: the origin (which, again, repels all trajectories except the separatrix and those on the second integral $n=0$ with $H_0 > 0$), a centre at $n^* = 41.49$ and a saddle at $n^* = 97.00$. The saddle is in the region of negative $\rho^*$. The physical trajectories are those with $H_0 > 0$ either to the right of the open part of the separatrix or between the stable curve and the unstable curve of the saddle. They diverge to $H \to \infty$ and $n \to \infty$. There are dynamically allowed  trajectories with $\rho_d < 0$ which exhibit cyclic behaviour.]
    {\label{F2c}\includegraphics[height=4.1cm,width=0.32\textwidth]{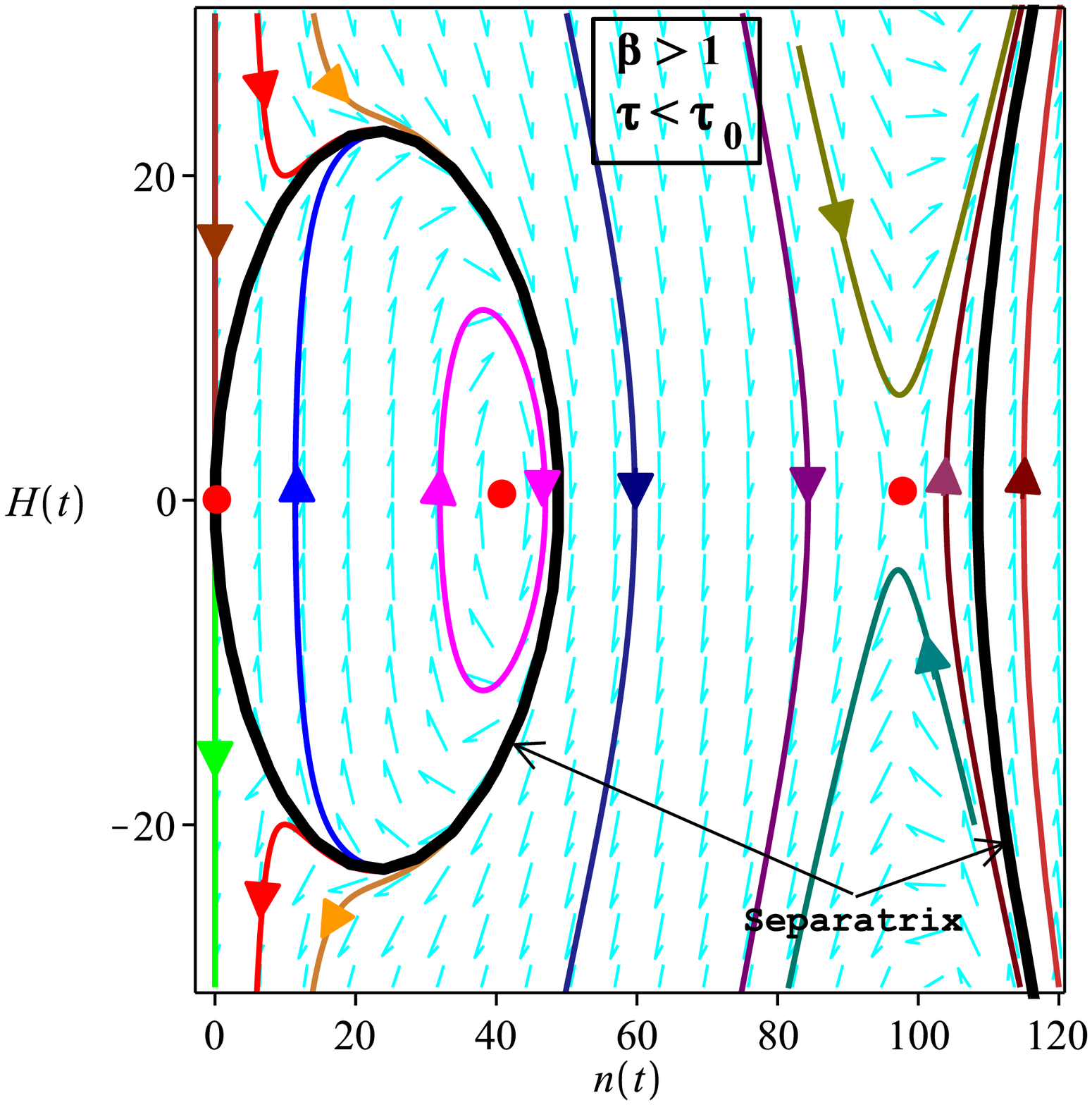}}
    \caption{\footnotesize{The case of $\beta > 1$.}}
    \label{Figure2}
    \end{figure}
    There is another critical point of the two-component dynamical system --- the origin $(n^{**} = 0, H^{**} = 0)$. The eigenvalues of the stability matrix are zero at the origin. \\
    To analyze the behaviour near the origin ($n \to 0,\, H \to 0$), expand the right-hand sides of the dynamical equations and keep only the leading
    terms in $n$ and $H$:
    \b
    \label{uno}
    \dot{n} & = & 3 (\beta - 1) n H, \\
    \label{hash0}
    \dot{H} & = & -\frac{3}{2} H^2 + \frac{\beta m_0}{2} n.
    \e
    There are two cases to consider. \\
    Firstly, when $\beta < 1$, then from $I_2(n, H) = C$ one has $3H^2 = m_0 n \,\,  + $ smaller terms. Thus (\ref{hash0}) becomes $\dot{H} = (1/2) (\beta - 1) m_0 n$. Introduce the Lyapunov function $G[n(t), H(t)] = n^2(t) + H^2(t)$. This function is strictly non-negative. Differentiating it with respect to time and substituting $\dot{n}$ and $\dot{H}$ with their corresponding expressions near the origin yields:
    \b
    \dot{G}[n(t), H(t)] = 2 n  \dot{n} + 2 H  \dot{H} = (\beta - 1) H  (6 n^2 + m_0 n).
    \e
    \begin{figure}[!ht]
    \centering
    \subfloat[\scriptsize When $\beta m_0 /(1 - 5 \beta / 2) >  (2 \beta B - B) / (\beta A - A)$,  $T^*(n^*)$  decreases monotonously from $\beta m_0 / (1 - 5 \beta / 2)$ at $n^* = 0$. The horizontal asymptote for the function $T^*(n^*)$ is $(2 \beta B - B) / (\beta A - A)$. There is one intersection point between the curves $T^*(n^*)$ and $T(n)$, irrespective of $\tau$. The critical points are the origin and a saddle see Figure 3c for $\tau > \tau_0$ and Figure 3d for $\tau < \tau_0$.]
    {\label{F3a}\includegraphics[height=4.3cm, width=0.32\textwidth]{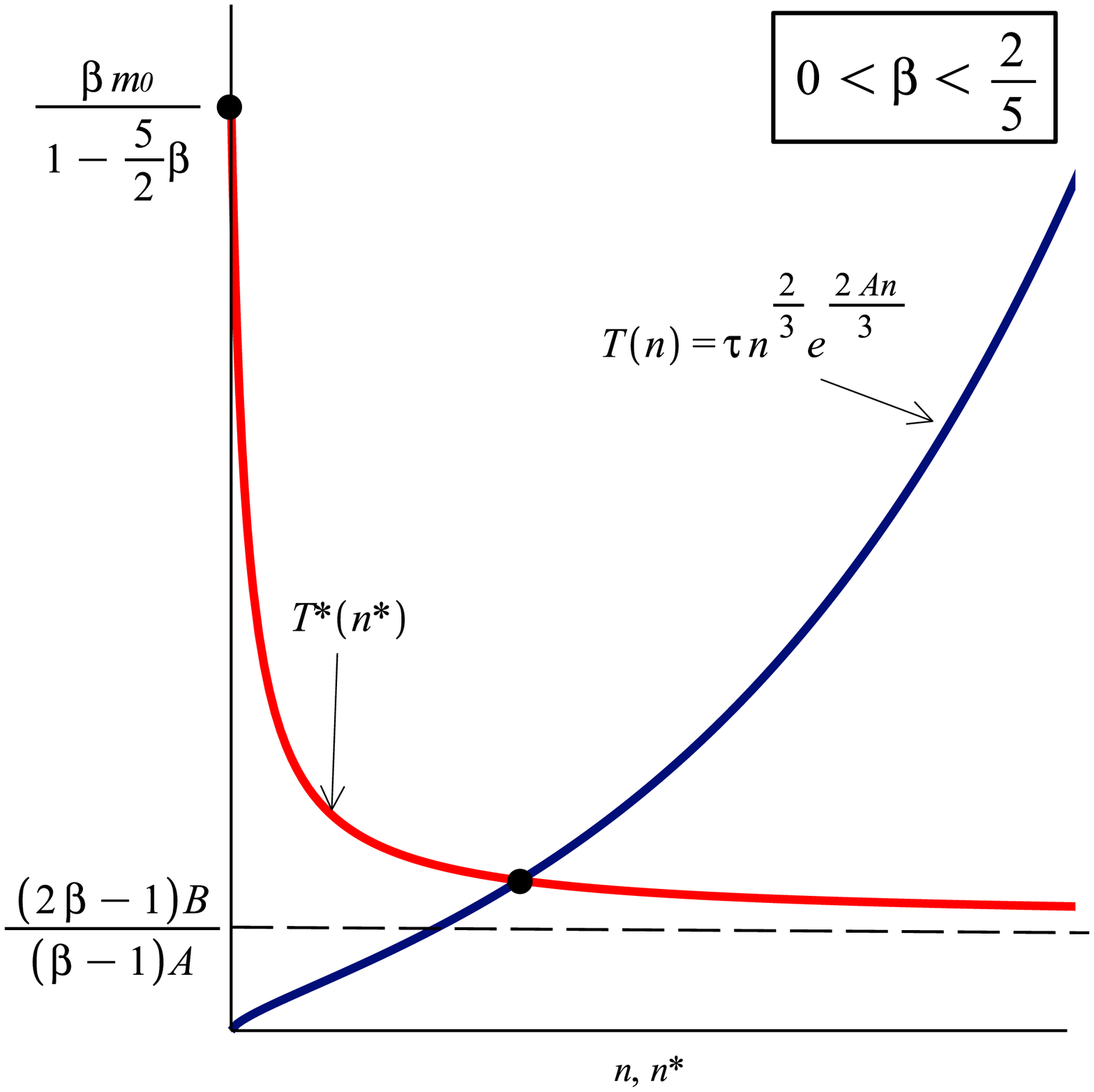}}
    \,
    \subfloat[\scriptsize The horizontal asymptote for the function $T^*(n^*)$ is again $(2 \beta B - B)/ (\beta A - A)$, but this time $\beta m_0 / (1 - 5 \beta / 2) <  (2 \beta B - B) / (\beta A - A)$. The function  $T^*(n^*)$  increases monotonously from $\beta m_0 / (1 - 5 \beta /2)$ at $n^* = 0$. There is either one intersection point between the curves $T^*(n^*)$ and $T(n)$ (depicted here) or three --- see Figure 3e for this case. The critical points here are, again, the origin and a saddle see Figure 3c fro $\tau > \tau_0$ and Figure 3d for $\tau < \tau_0$. ]
    {\label{F3b}\includegraphics[height=4.3cm,width=0.32\textwidth]{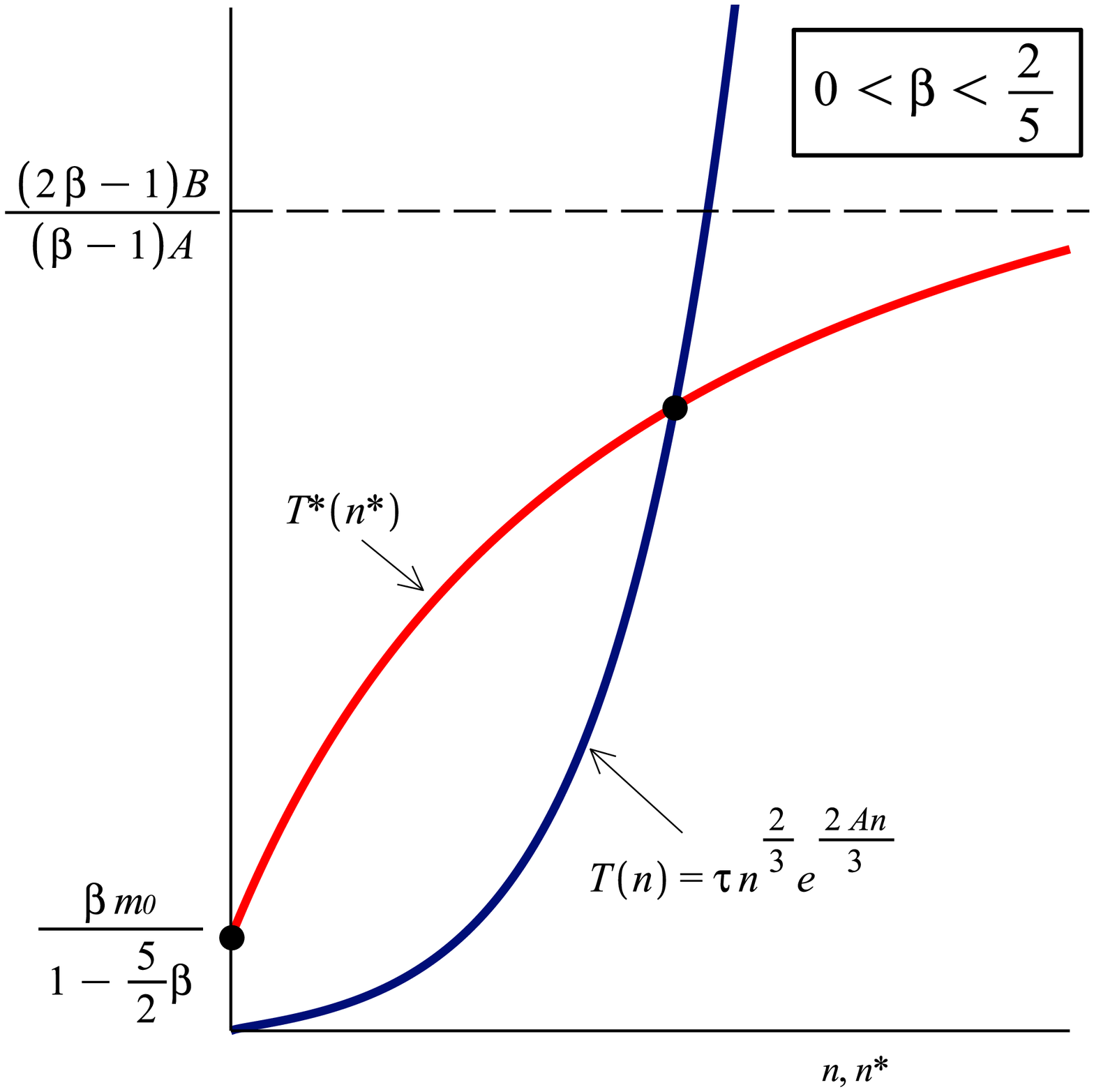}}
    \,
    \subfloat[\scriptsize When $\tau = 15$ (i.e. $\tau > \tau_0$) and $\beta = 0.39$ (that is, $0 < \beta <  2/5$), the situation on Figure 3a applies. The critical points are the origin and a saddle at $n^* = 72.76$. The physical trajectories are those with $H_0 > 0$ which are to the left of the stable curve of the saddle. They all converge to the origin in infinite time. ]
    {\label{F3c}\includegraphics[height=4.3cm,width=0.32\textwidth]{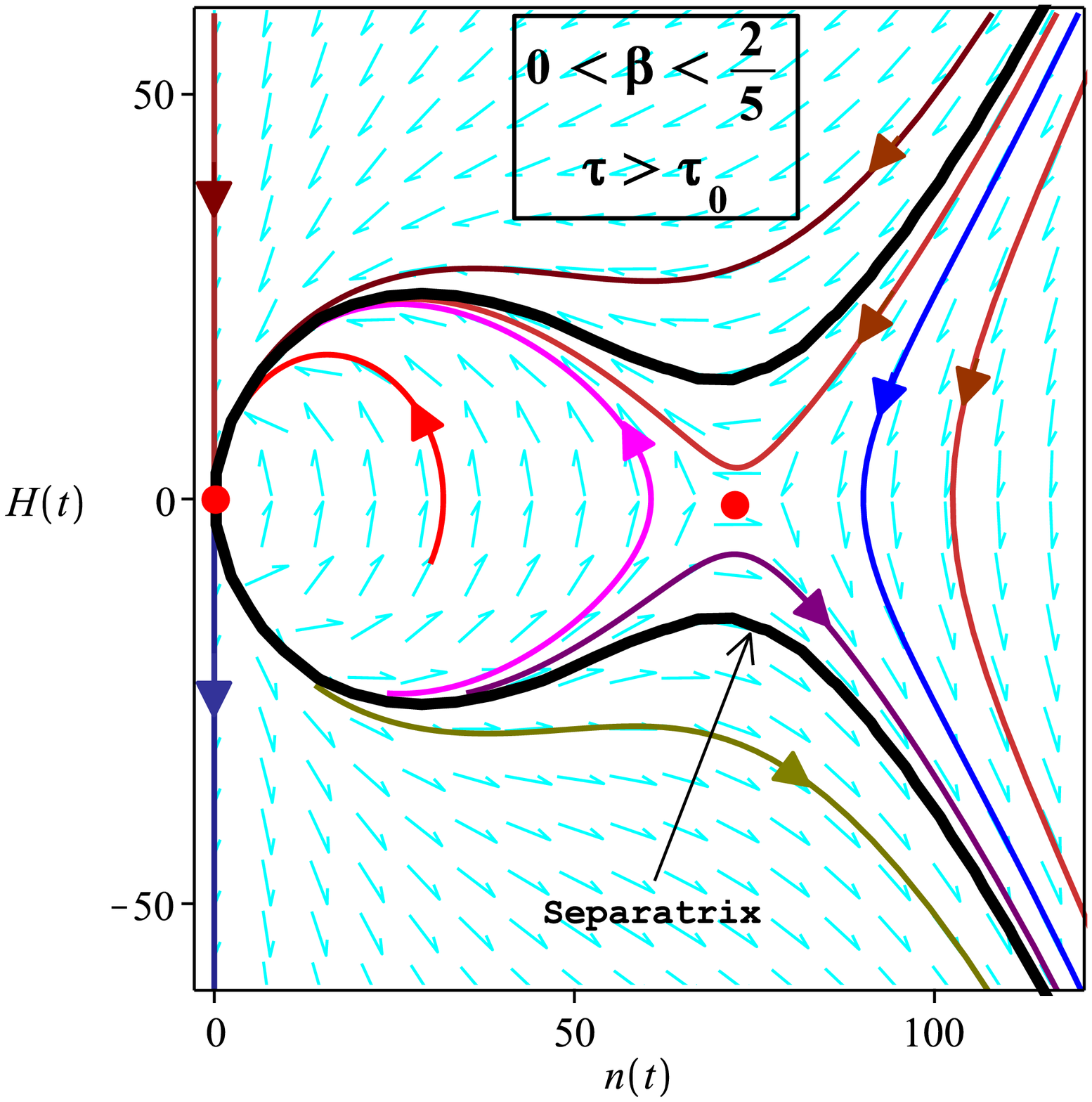}}
    \caption{\footnotesize Parts (a) to (c) --- the case of $0 < \beta < \frac{1}{2}$. }
    \label{Figure3_1}
    \end{figure}

    \addtocounter{figure}{-1}
    \addtocounter{subfigure}{+3}

    \begin{figure}[!ht]
    \centering

    \subfloat[\scriptsize  When $\tau = 14$ (that is, $\tau < \tau_0$) and $\beta = 0.39$ (that is, $0 < \beta <  2/5$), the situation on Figure 3a applies again. There are, again, two critical points --- the origin and a saddle at $n^* = 76.71$. The physical trajectories are those with $H_0 > 0$ to the left of the stable curve of the saddle, including those with $\rho_d < 0$ which are inside the closed loop of the separatrix. The trajectories converge to the origin in infinite time. At the saddle point, $\rho^*$ is negative. Homoclinic orbits are present.]
    {\label{F3d}\includegraphics[height=4.3cm,width=0.32\textwidth]{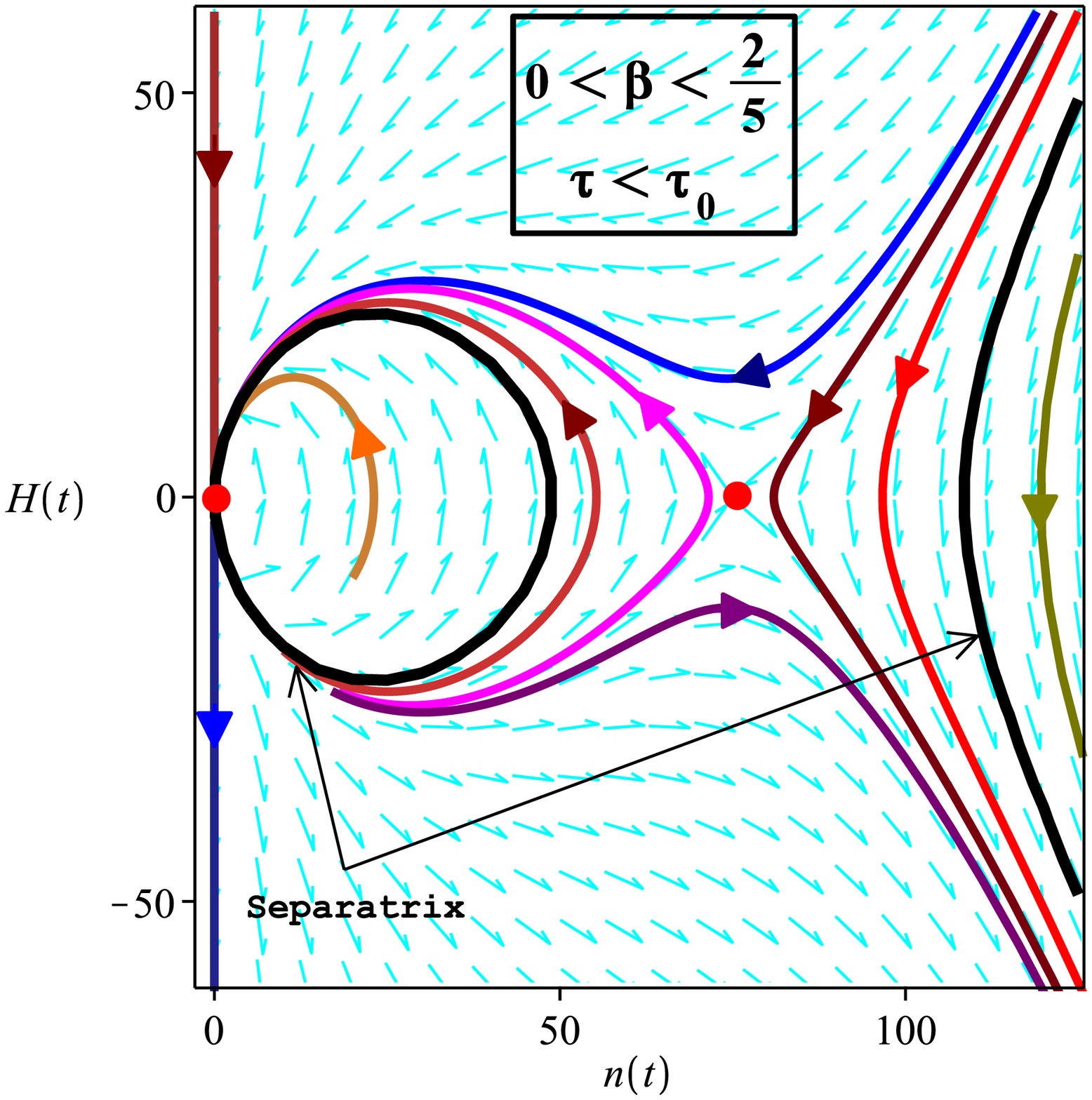}}
    \,
    \subfloat[\scriptsize When $\beta m_0 / (1 - 5 \beta / 2) <  (2 \beta B - B) / (\beta A - A)$, both curves $T^*(n^*)$ and $T(n)$ increase monotonously and, depending on $\tau$ and $\beta$, there may be one intersection point between them (see Figure 3b for this case) or there may be three intersection points between them. In the latter case, there are four critical points --- the origin, a saddle, a centre, and another saddle, in order of increasing $n^*$ --- see Figures 3f, 3g, 3h, and 3i.]
    {\label{F3e}\includegraphics[height=4.3cm,width=0.32\textwidth]{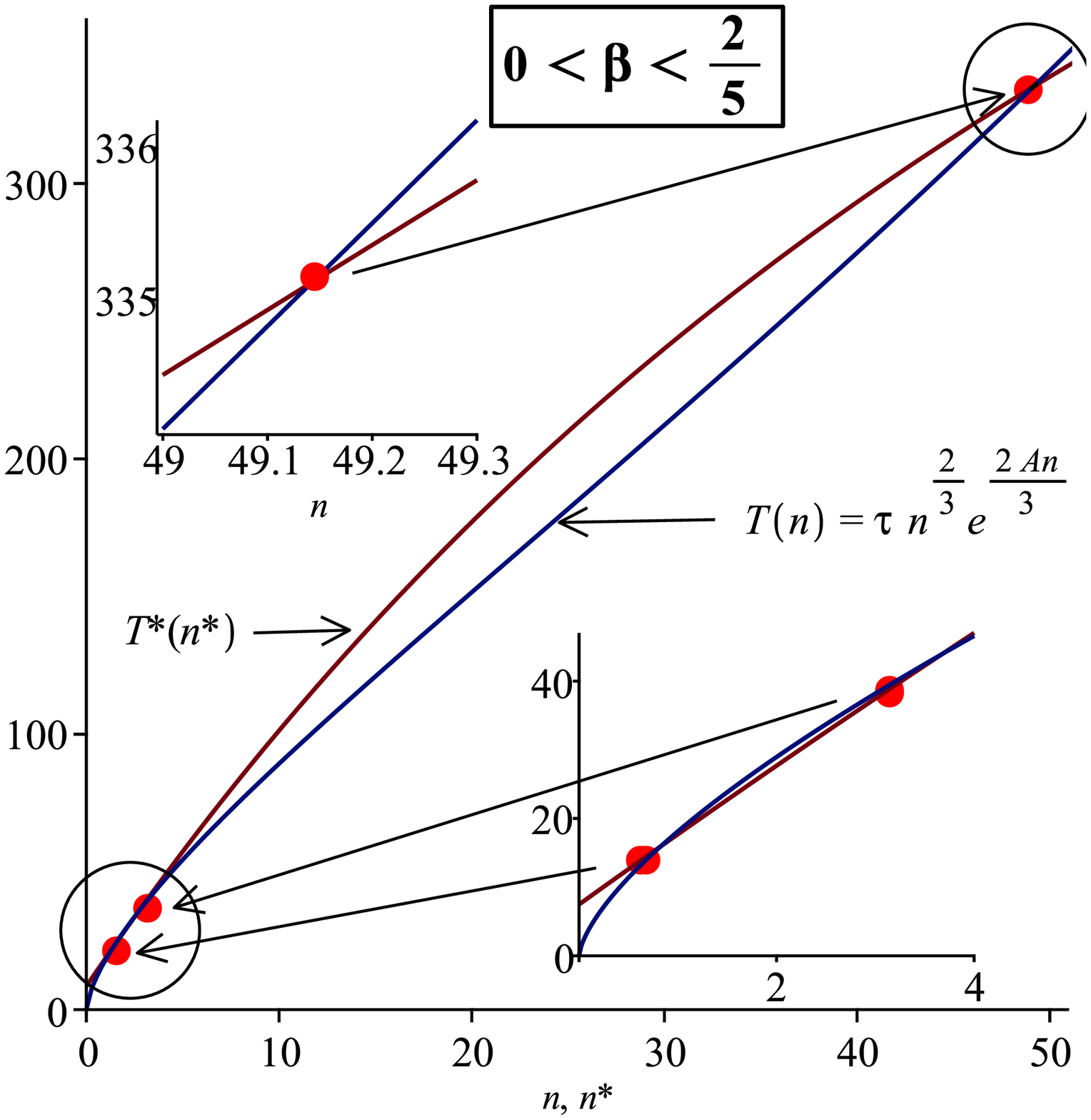}}
    \,
    \subfloat[\scriptsize When $\tau = 18$ (that is, $\tau > \tau_0$) and $\beta = 0.07$ (that is, $0 < \beta <  2/5$), the situation on Figure 3e applies. There are four critical points: the origin and the three intersections of the curves $T^*(n^*)$ and $T(n)$: the saddle at $n^* = 1.38$, the centre at $n^*=2.68$ (all shown here) and the saddle at $n^*=49.14$ which is shown on Figure 3g. The physical trajectories are those with $H_0 > 0$ which are to the left of the stable curve of the saddle at $n^*=49.14$ --- drawn on Figure 3g which shows the region of higher number densities. The trajectories converge to the origin in infinite time. Again, there are dynamically allowed trajectories with cyclic behaviour.]
    {\label{F3f}\includegraphics[height=4.3cm,width=0.32\textwidth]{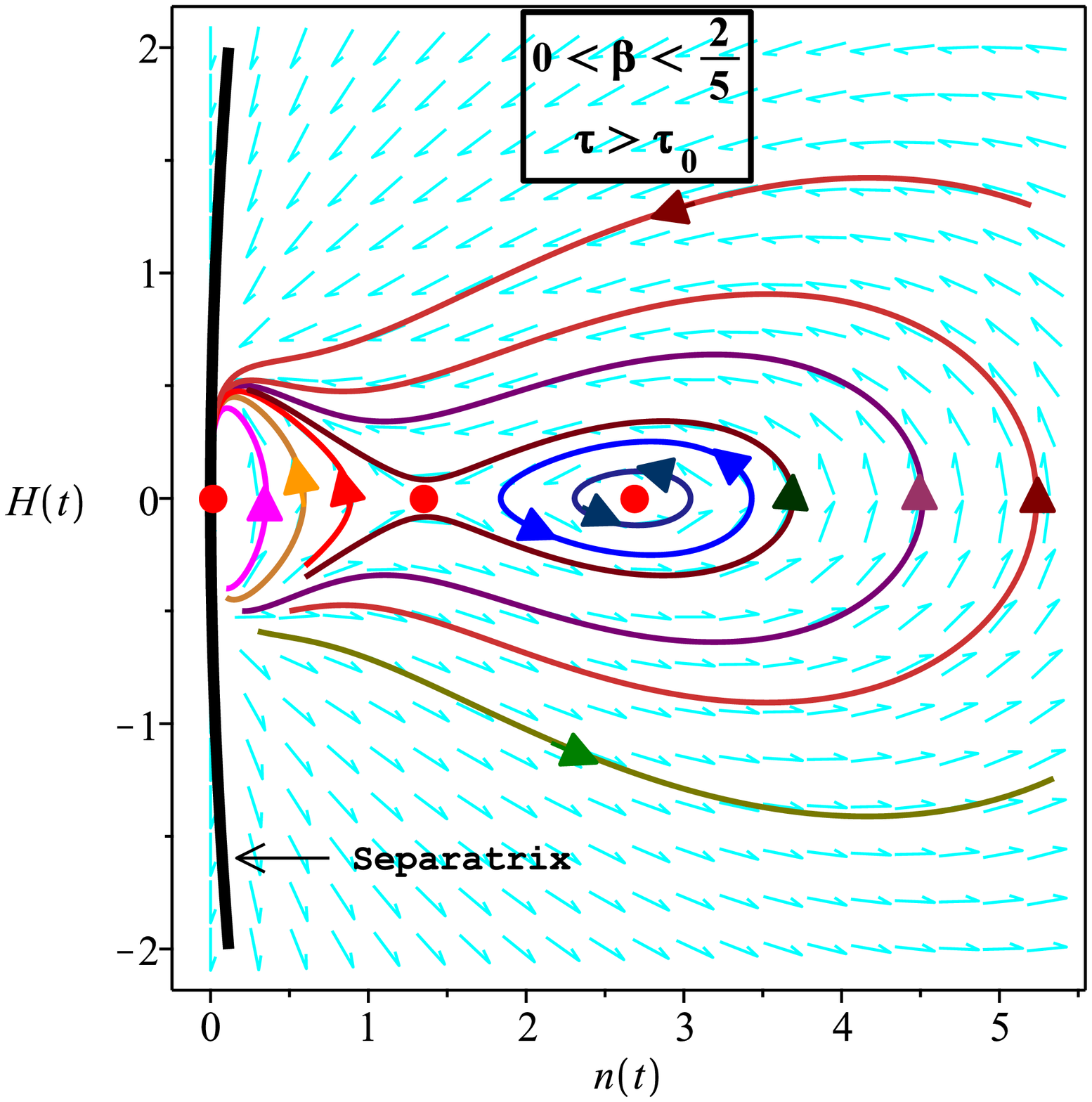}}
    \caption{\footnotesize{Parts (d) to (f) --- the case of $0 < \beta < \frac{1}{2}$.}}
    \label{Figure3_2}
    \end{figure}

    \addtocounter{figure}{-1}
    \addtocounter{subfigure}{+6}

    \begin{figure}[!ht]
    \centering
    \subfloat[\scriptsize Continuation of Figure 3f for the region of higher number densities for the case of $\tau = 18$ (that is, $\tau > \tau_0$) and $\beta = 0.07$ (that is, $0 < \beta <  2/5$). The situation on Figure 3e applies. There are four critical points: the origin and the three intersections of the curves $T^*(n^*)$ and $T(n)$: the saddle at $n^* = 1.38$, the centre at $n^*=2.68$ (all shown on Figure 3f) and the saddle at $n^*=49.14$ shown here. The physical trajectories are those with $H_0 > 0$ which are to the left of the stable curve of the saddle at $n^*=49.14$ and they all converge to the origin in infinite time. ]
    {\label{F3g}\includegraphics[height=4.3cm,width=0.32\textwidth]{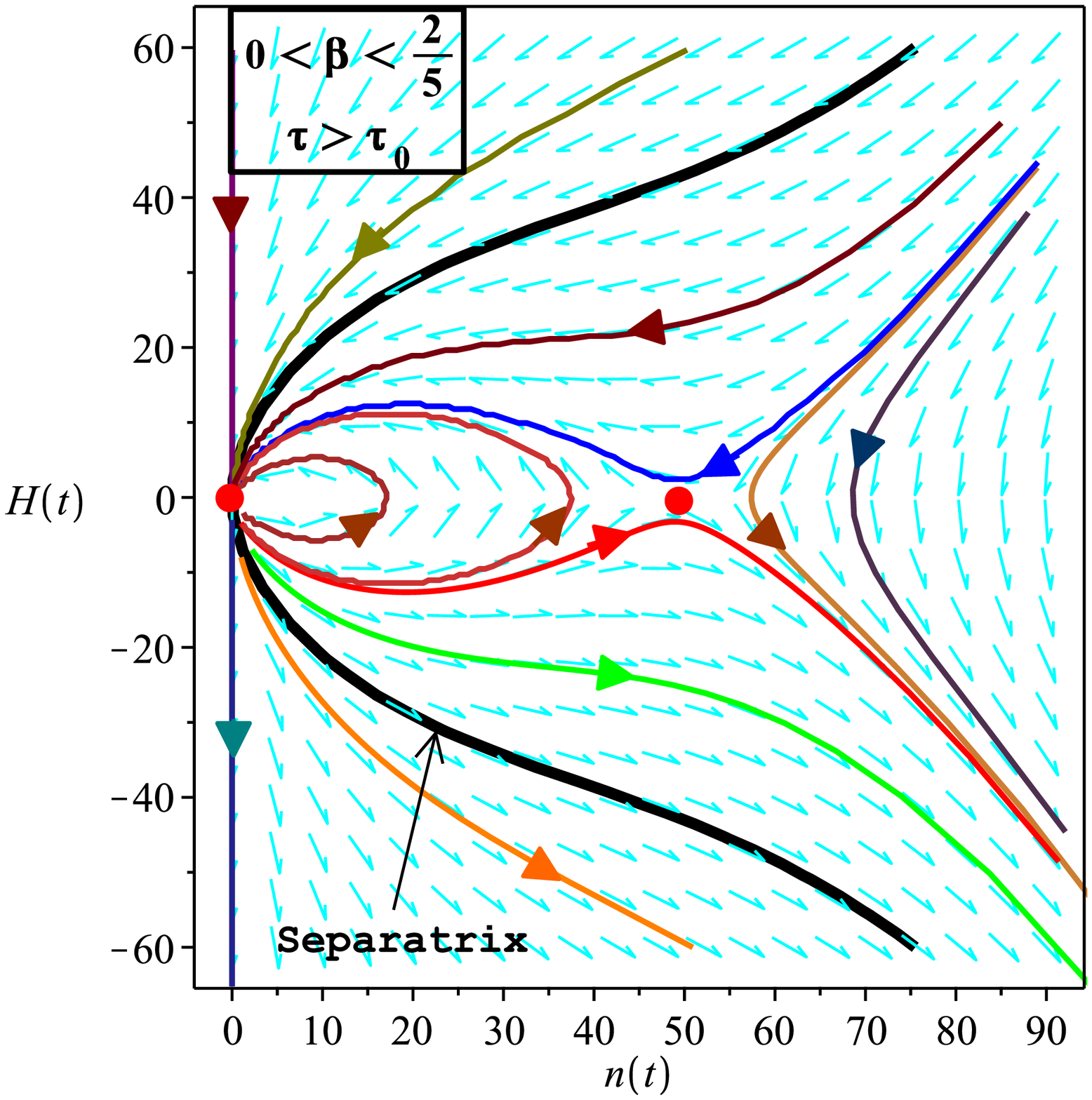}}
    \,
    \subfloat[\scriptsize
    When $\tau = 14.5$ (that is, $\tau < \tau_0$) and $\beta = 0.035$ (that is, $0 < \beta <  2/5$), the situation on Figure 3e applies again. There are four critical points: the origin and the three intersections of the curves $T^*(n^*)$ and $T(n)$: the saddle at $n^* = 0.40$, the centre at $n^*=1.77$ (all shown here) and the saddle at $n^* = 75.93$ which is shown on Figure 3i. The physical trajectories are those with $r_d > 0$ and $H_0 > 0$ which are to the left of the stable curve of the saddle at $n^* = 75.93$ (drawn on Figure 3i which shows the region of higher number densities), or the trajectories inside the closed loop of the separatrix which are with $H_0 > 0$ and to the left of the stable curve of the saddle at $n^* = 0.40$, drawn here.
    The physical trajectories converge to the origin in infinite time.]
    {\label{F3h}\includegraphics[height=4.3cm,width=0.32\textwidth]{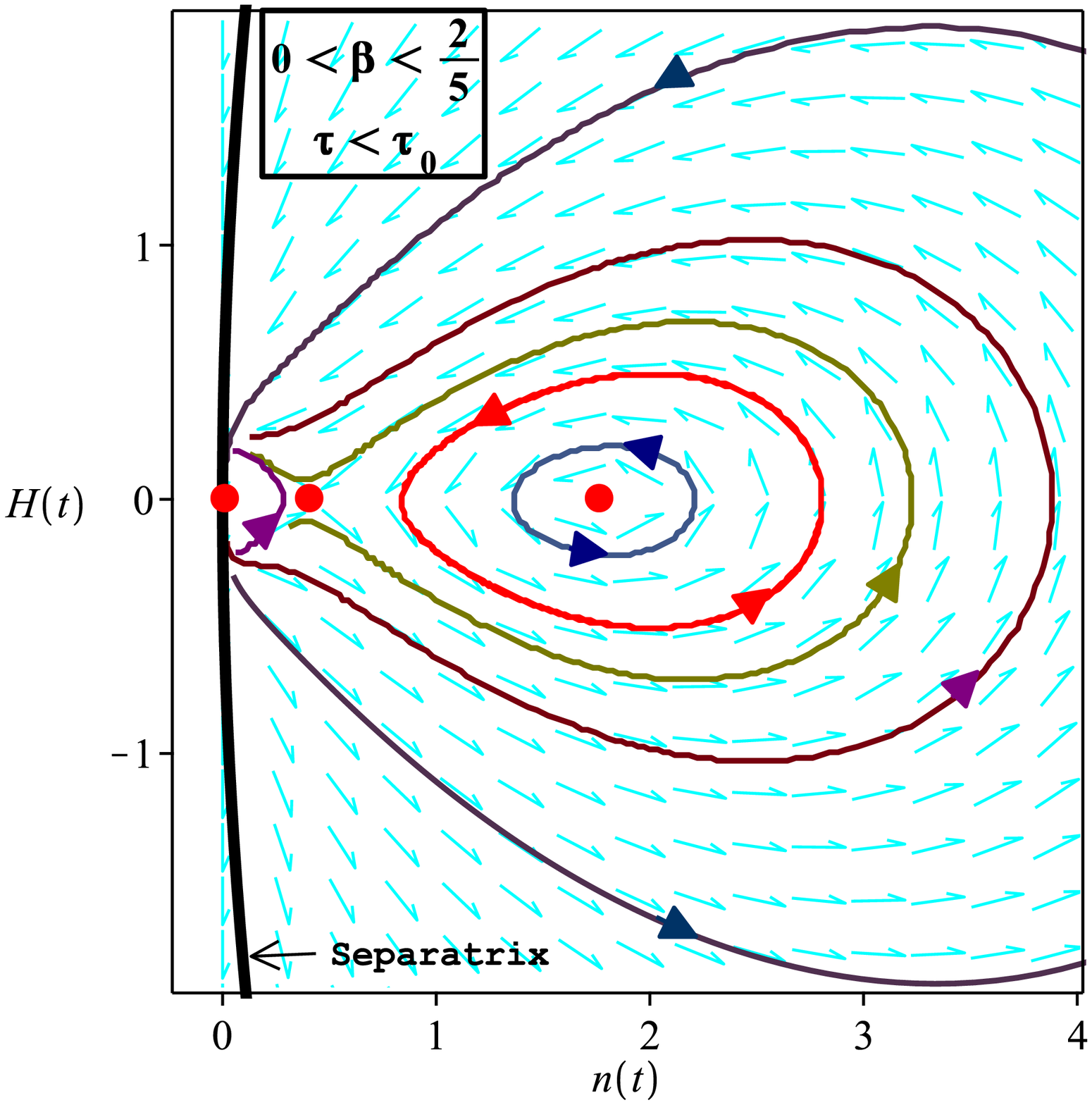}}
    \,
    \subfloat[\scriptsize Continuation of Figure 3h for the region of higher number densities for the case of $\tau = 14.5$ (that is, $\tau < \tau_0$) and $\beta = 0.035$ (that is, $0 < \beta <  2/5$). The situation on Figure 3e applies again. There are four critical points: the origin and the three intersections of the curves $T^*(n^*)$ and $T(n)$: the saddle at $n^* = 0.40$, the centre at $n^*=1.77$ (all shown on Figure 3h) and the saddle at $n^* = 75.93$ which is shown here. The physical trajectories are those with $r_d > 0$ and $H_0 > 0$ which are to the left of the stable curve of the saddle at $n^* = 75.93$ (shown here), or the trajectories inside the closed loop of the separatrix which are with $H_0 > 0$ and to the left of the stable curve of the saddle at $n^* = 0.40$, shown on Figure 3h which shows the region of lower number densities. The physical trajectories converge to the origin in infinite time.]
    {\label{F3i}\includegraphics[height=4.3cm,width=0.32\textwidth]{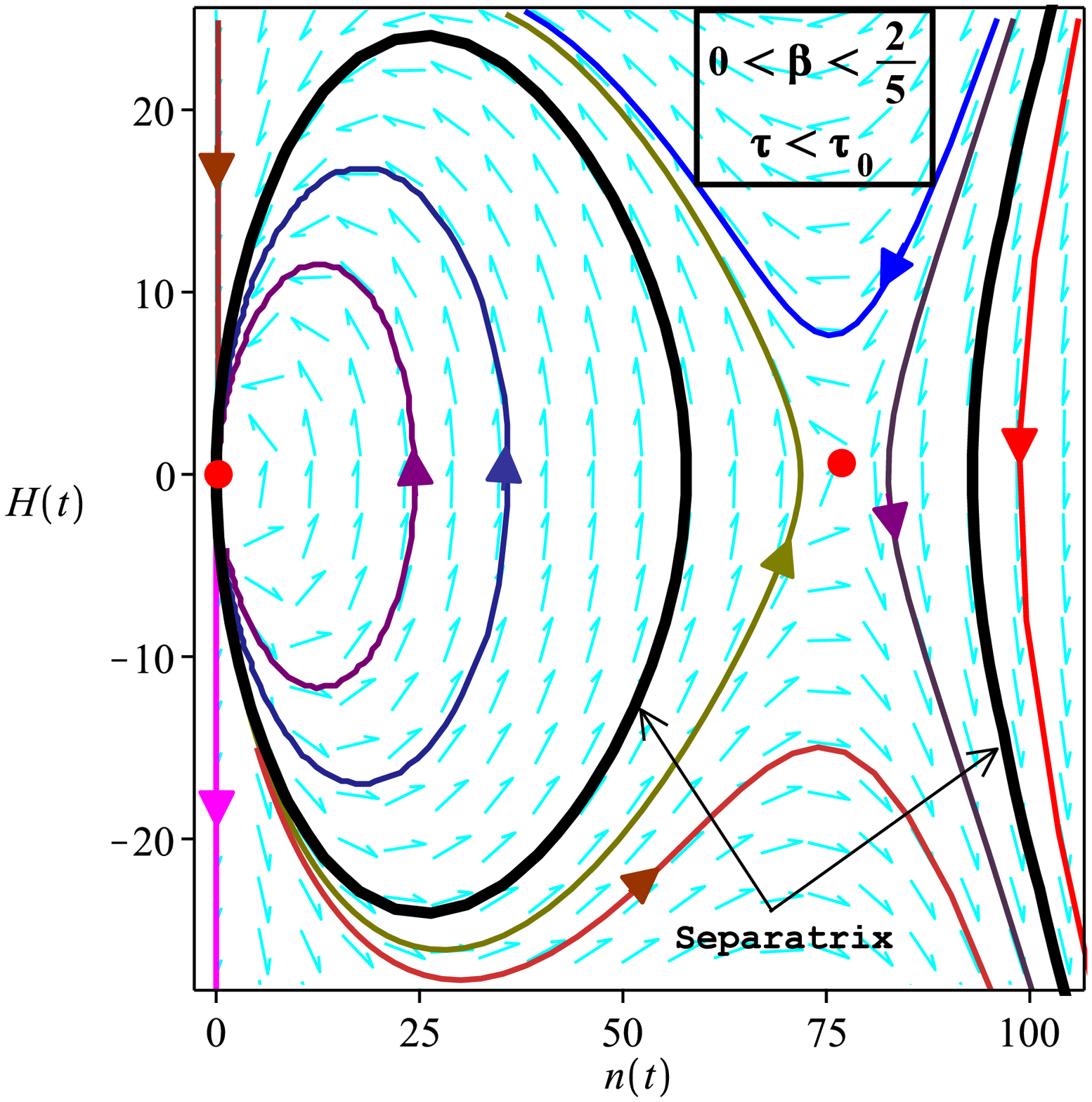}}
    \caption{\footnotesize Parts (g) to (i) --- the case of $0 < \beta < \frac{1}{2}$. }
    \label{Figure3_3}
    \end{figure}
\n
This is negative in the upper half-plane $H > 0$ (thus the origin attracts trajectories from the upper half-plane) and positive in the lower half plane (thus trajectories in the lower half-plane $H < 0$ are repelled by the origin). \\
    As an alternative point of view when $\beta < 1$, one can consider the trajectories near the origin (including the separatrix) and obtain the asymptotic behaviour of $n(t)$ as $t \to \infty$. Using $3 H^2 =  m_0 n \,\, + $ smaller terms in  $\dot{n} = 3 (\beta -1) H n$ gives:
    \b
    \label{bigfreeze1}
    n(t) = \frac{n_0}{\bigl[ 1 + \frac{1}{2} \sigma (1 - \beta) \sqrt{3 m_0 n_0}(t-t_0)\Bigr]^2},
    \e
    where $\sigma = +1$ for trajectories in the half-plane $H > 0$ and $\sigma = -1$ for those in the half-plane $H < 0$. For the trajectories in the upper half-plane, one has $n(t) \simeq 1/t^2$, while for those in the lower half-plane, $n(t)$ increases with time. \\
    One also has $\dot{H} = (3/2)(\beta-1) H^2$ or
    \b
    \label{bigfreeze2}
    H(t) =  \frac{H_0}{1 + \frac{3}{2}(1 - \beta) H_0 (t - t_0)}.
    \e
    Therefore $H$ decays to zero ($H \simeq 1/t$) for trajectories in the upper half-plane or $H$ decreases with time for trajectories in the lower half-plane. \\
    Clearly, the origin is reachable in infinite time along the trajectories in the $H > 0$ half-plane. \\
    Secondly, when $\beta > 1$, one can look at the separatrix $3H^2 - n[m_0 + (3/2)T] + Bn^2 = 0$. As discussed, this curve is a second integral and it passes through the origin. Along the separatrix near the origin, one has $3H^2 = m_0 n \,\,  + $ smaller terms and, along the separatrix only, one also has $\dot{n} = 3 (\beta - 1) n H$ and $\dot{H} = (3/2)(\beta - 1) H^2$ near the origin. The solutions to these two equations are given by (\ref{bigfreeze1}) and (\ref{bigfreeze2}), respectively. The difference between the cases $\beta < 1$ and the current case $\beta > 1$ lies in the fact that the solutions to (\ref{bigfreeze1}) and (\ref{bigfreeze2}) apply to all trajectories near the origin when $\beta < 1$, while (\ref{bigfreeze1}) and (\ref{bigfreeze2}) apply only to the separatrix when $\beta > 1$. It is now obvious that the separatrix enters the origin from the lower half-plane $H < 0$ and exits it from the upper half-plane $H > 0$. Also, it takes an infinite amount of time to enter the origin. \\

    \begin{figure}[!ht]
    \centering
    \subfloat[\scriptsize When $\beta m_0 / (2 \beta B - B) < (1 - 5 \beta / 2)  / (\beta A - A)$, the curve $T^*(n^*)$ increases monotonously from $\beta m_0 / (1 - 5 \beta / 2) < 0$ to infinity at $(1 - 5 \beta / 2) / (\beta A - A )$. Positive values of $T^*(n^*)$ exist for $n^*$ between  $\beta m_0 / (2 \beta B - B)$  and the vertical asymptote $(1 - 5 \beta / 2)  / (\beta A - A)$. Drawn here is the case of one intersection point between $T^*(n^*)$ and $T(n)$ --- see Figure 4e for the case of three intersection points. In the case of one intersection point, the critical points are the origin and a saddle.  The trajectories are on Figure 4c for $\tau > \tau_0$ and Figure 4d for $\tau < \tau_0$.]
    {\label{F4a}\includegraphics[height=5.2cm, width=0.49\textwidth]{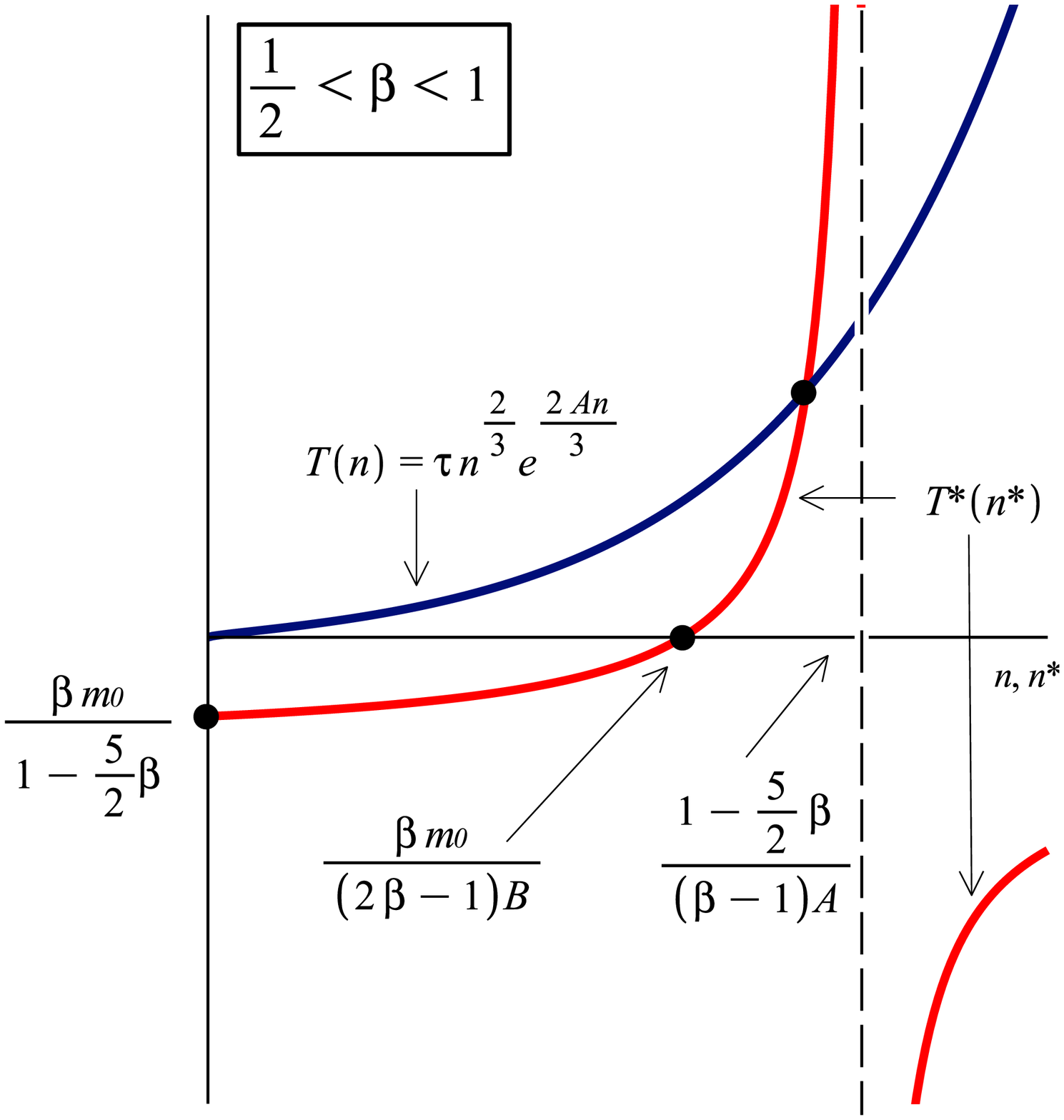}}
    \quad
    \subfloat[\scriptsize When $\beta m_0 / (2 \beta B - B) > (1 - 5 \beta / 2)  / (\beta A - A)$, positive values of the monotonously decreasing function $T^*(n^*)$ are between the vertical asymptote $(1 - 5 \beta / 2)  / (\beta A - A)$ and $\beta m_0 / (2 \beta B - B).$
    There is always one intersection point between $T^*(n^*)$ and $T(n)$ and the critical points are, again, the origin and a saddle. The trajectories are on Figure 4c for $\tau > \tau_0$ and Figure 4d for $\tau < \tau_0$.]
    {\label{F4b}\includegraphics[height=5.2cm,width=0.48\textwidth]{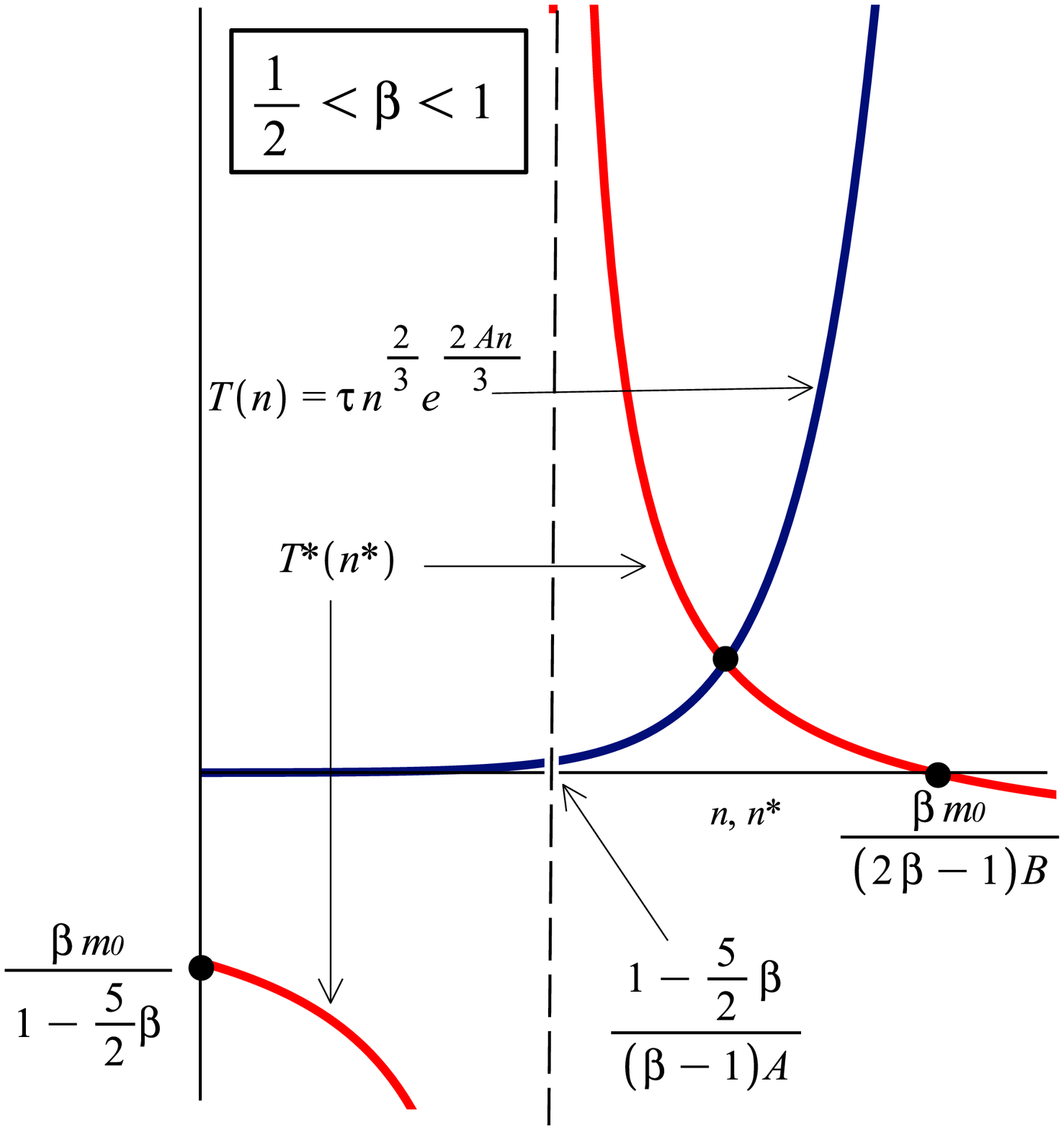}}
    \caption{\footnotesize{Parts (a) to (b) --- the case of
    $\frac{1}{2} < \beta < 1$.}}
    \label{Figure4_1}
    \end{figure}

    \addtocounter{figure}{-1}
    \addtocounter{subfigure}{+2}

    \begin{figure}[!ht]
    \centering
    \subfloat[\scriptsize When $\tau = 16$ (that is, $\tau > \tau_0$) and $\beta = 0.55$ (that is, $1/2 < \beta <  1$), one has $\beta m_0 / (2 \beta B - B) < (1 - 5 \beta / 2)  / (\beta A - A)$. Thus, the situation on Figure 4a applies. There are two critical points: the origin and the only intersection point of the curves $T^*(n^*)$ and $T(n)$ --- the saddle at $n^* = 74.14$. The physical trajectories are those with $H_0 > 0$ which are to the left of the stable curve of the saddle. They converge to the origin in infinite time.]
    {\label{F4c}\includegraphics[height=4.3cm,width=0.32\textwidth]{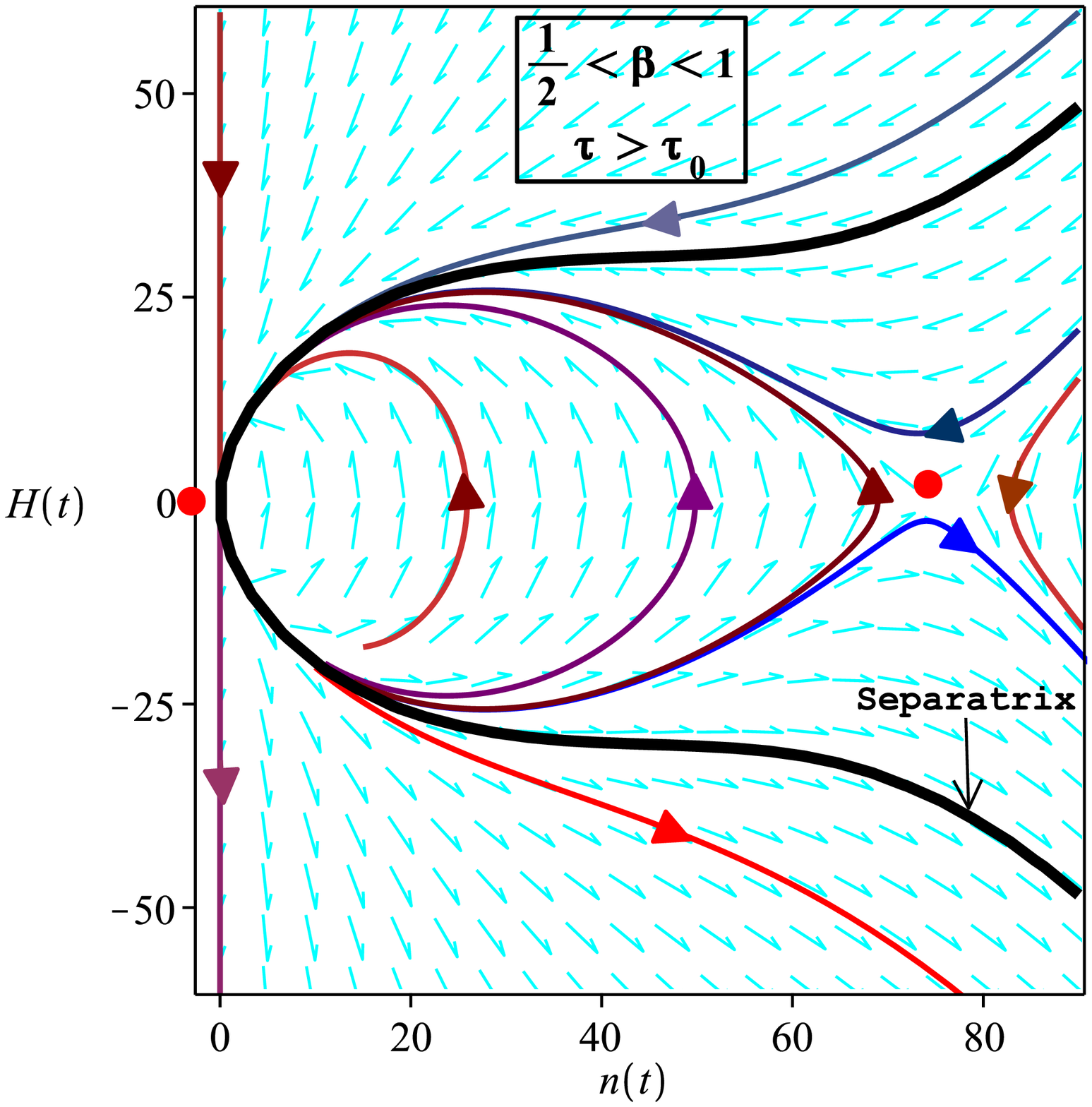}}
    \,
    \subfloat[\scriptsize When $\tau = 14$ (that is, $\tau > \tau_0$) and $\beta = 0.55$ (that is, $1/2 < \beta <  1$), one again has $\beta m_0 / (2 \beta B - B) < (1 - 5 \beta / 2)  / (\beta A - A)$. Thus, the situation on Figure 4a applies again. There are two critical points: the origin and the only intersection point of the curves $T^*(n^*)$ and $T(n)$ --- the saddle at $n^* = 73.20$. The physical trajectories are those with $H_0 > 0$ which are to the left of the stable curve of the saddle, including the ones with $\rho_d < 0$ which are inside the closed loop of the separatrix. The physical trajectories converge to the origin in infinite time.]
    {\label{F4d}\includegraphics[height=4.3cm,width=0.32\textwidth]{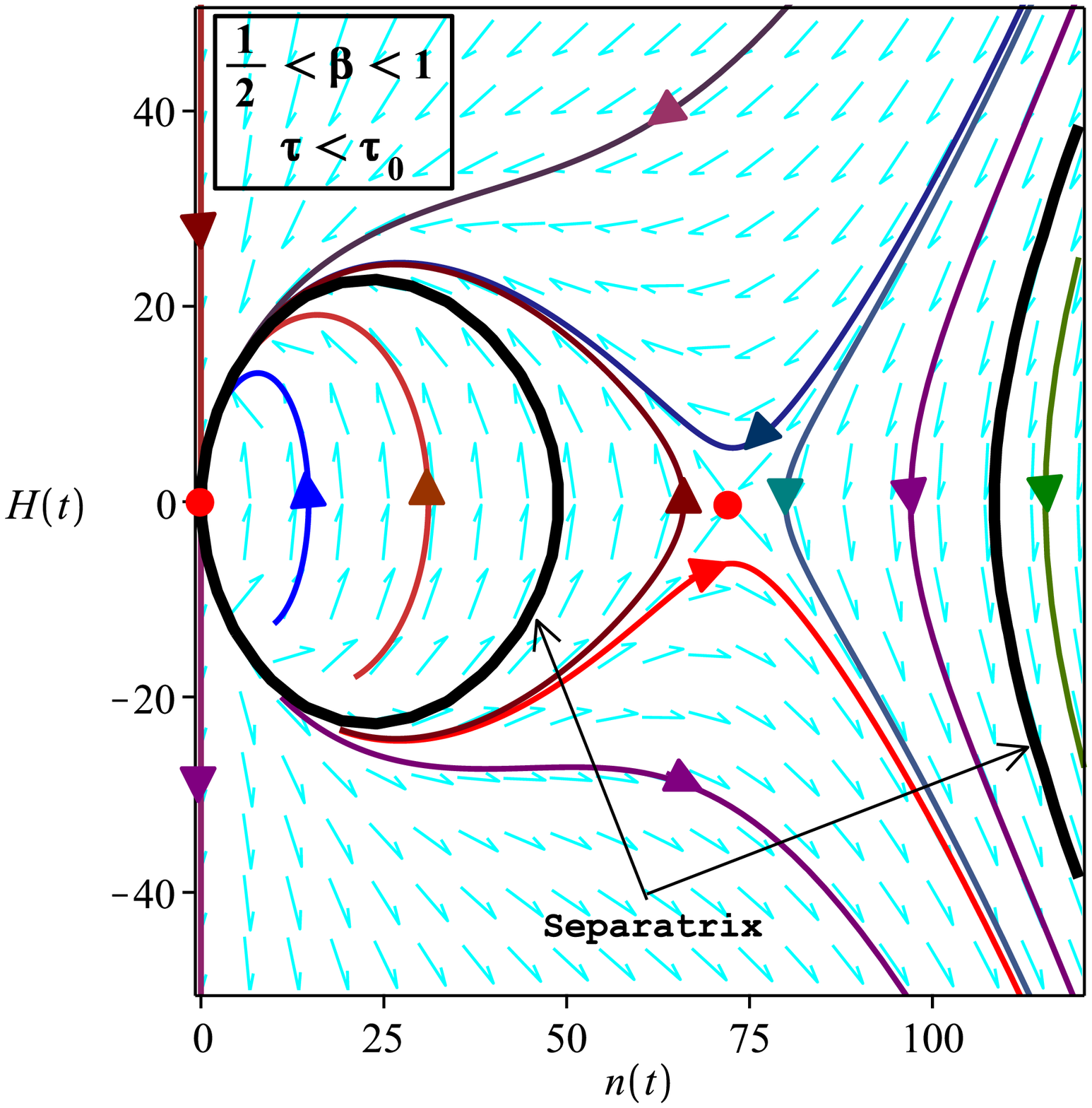}}
    \,
    \subfloat[\scriptsize  When $\beta m_0 / (2 \beta B - B) < (1 - 5 \beta / 2)  / (\beta A - A)$, both curves $T^*(n^*)$ and $T(n)$ increase monotonously and, depending on $\tau$ and $\beta$, there may be a case of three intersection points between them --- as illustrated here. There are four critical points --- the origin, a saddle, a centre, and another saddle, in order of increasing $n^*$ --- see Figures 4f and 4g.]
    {\label{F4e}\includegraphics[height=4.3cm,width=0.32\textwidth]{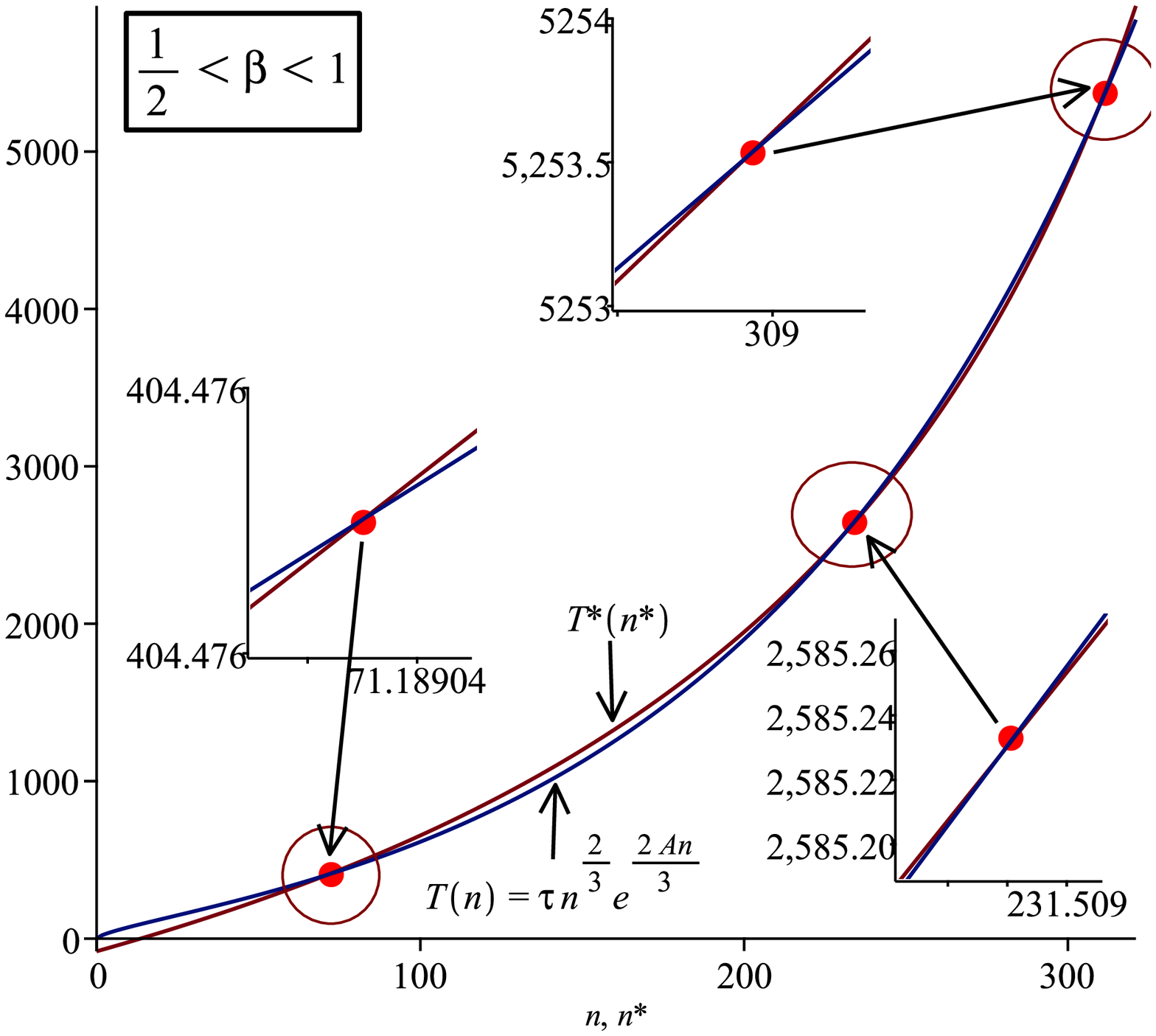}}
    \caption{\footnotesize{Parts (c) to (e) --- the case of
    $\frac{1}{2} < \beta < 1$.}}
    \label{Figure4_2}
    \end{figure}

    \addtocounter{figure}{-1}
    \addtocounter{subfigure}{+5}

    \begin{figure}[htbp]
    \centering

    \subfloat[\scriptsize When $\tau = 15$ (that is, $\tau > \tau_0$) and $\beta = 0.79$ (that is, $1/2 < \beta <  1$), one has $\beta m_0 / (2 \beta B - B) < (1 - 5 \beta / 2)  / (\beta A - A)$. The situation on Figure 4e applies. There are four critical points: the origin and the three intersection points of the curves $T^*(n^*)$ and $T(n)$: the saddle at $n^* = 78.21$, the centre at $n^* = 200.98$ and the saddle at $n^* = 326.43$. The physical trajectories are those with $H_0 > 0$ which are to the left of the stable curve of the saddle at $n^* = 78.21$. They converge to the origin in infinite time. Again, there are dynamically allowed periodic trajectories.]
    {\label{F4f}\includegraphics[height=4.3cm,width=0.32\textwidth]{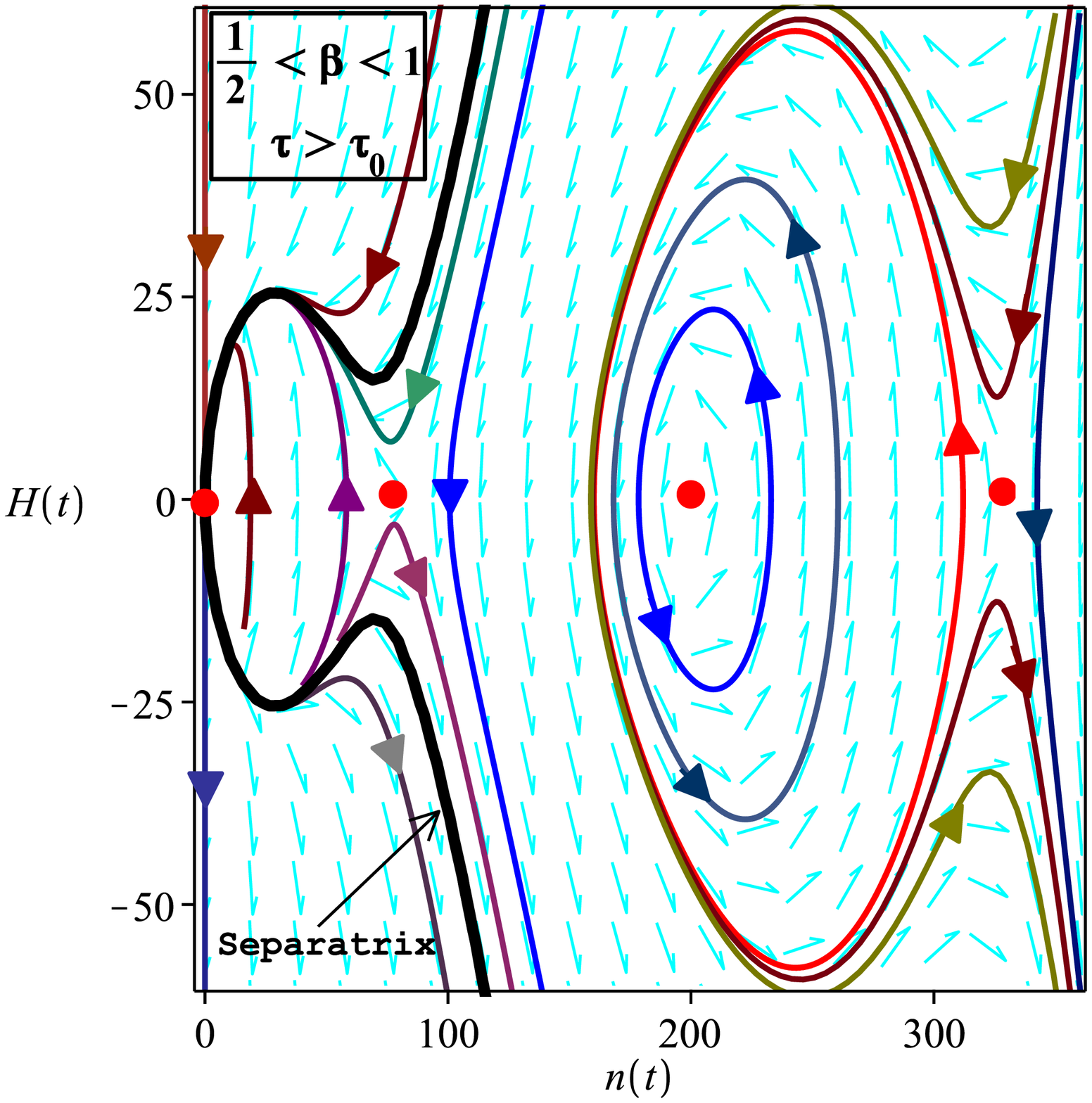}}
    \,
    \subfloat[\scriptsize
    When $\tau = 14.65$ (that is, $\tau < \tau_0$) and $\beta = 0.79$ (that is, $1/2 < \beta <  1$), one again has $\beta m_0 / (2 \beta B - B) < (1 - 5 \beta / 2)  / (\beta A - A)$. The situation on Figure 4e applies again. There are four critical points: the origin and the three intersection points of the curves $T^*(n^*)$ and $T(n)$: the saddle at $n^* = 71.19$, the centre at $n^* = 231.51$ and the saddle at $n^* = 309.00$. The physical trajectories are those with $H_0 > 0$ which are to the left of the stable curve of the saddle at $n^* = 71.19$. They converge to the origin in infinite time. Again, there are dynamically allowed periodic trajectories.]
    {\label{F4g}\includegraphics[height=4.3cm, width=0.32\textwidth]{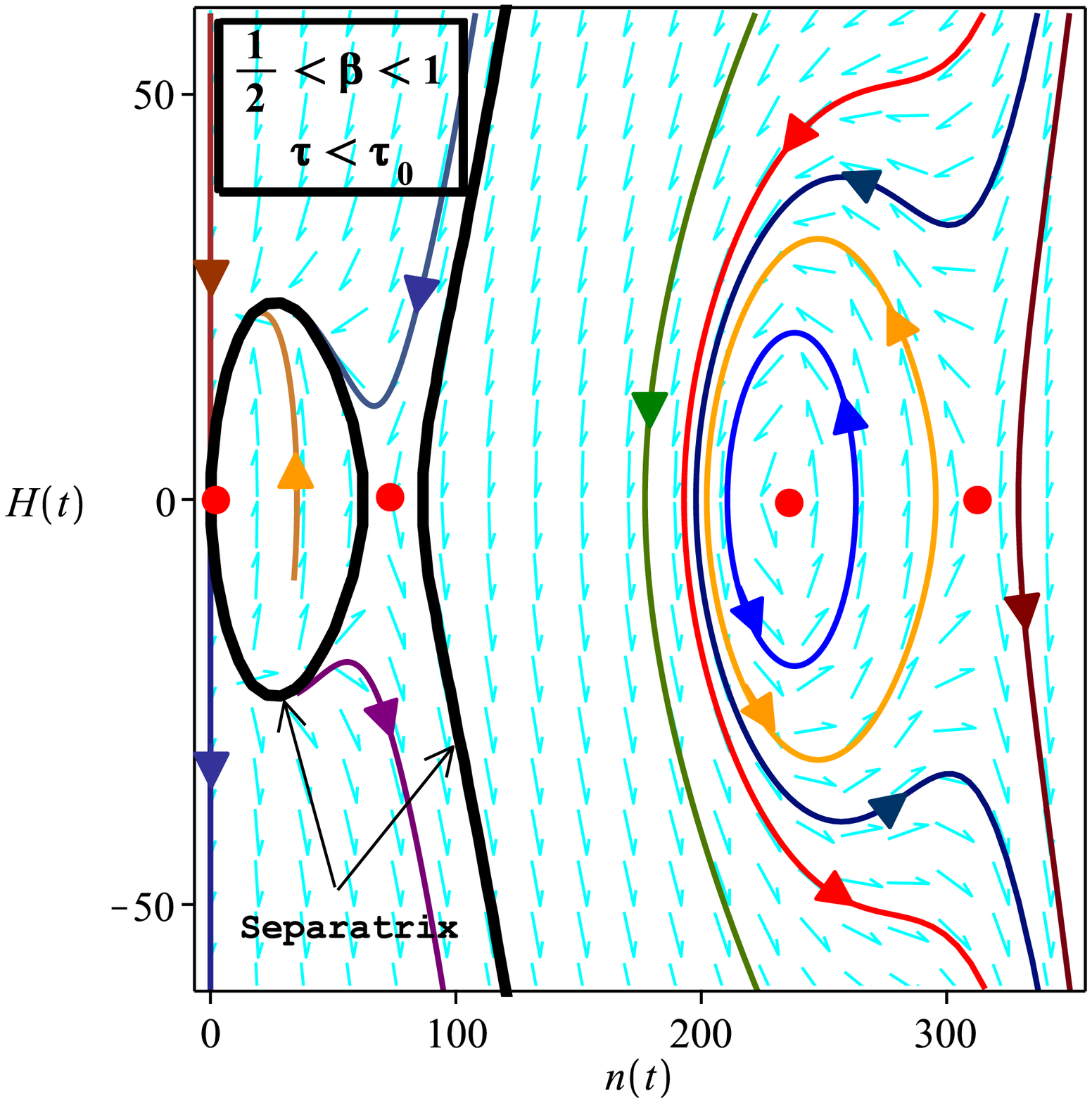}}
    \,
    \subfloat[\scriptsize The separate case of $\frac{2}{5} < \beta < \frac{1}{2}$ is included here due to the similarities with the situation on Figure 4b. Again, $(1 - 5 \beta / 2)  / (\beta A - A)$ is a vertical asymptote and the positive values of the monotonously decreasing function $T^*(n^*)$ are to the right of it. The difference between this case and the one on Figure 4b is in the presence of a horizontal asymptote at $(2 \beta B - B)  / (\beta A - A)$.
    There is always one intersection point between $T^*(n^*)$ and $T(n)$ --- at $n^* > (1 - 5 \beta / 2)  / (\beta A - A)$  --- and the critical points are, again, the origin and a saddle. The trajectories are as those on Figure 4c for $\tau > \tau_0$ and as those on Figure 4d for $\tau < \tau_0$ (in the latter case $\rho^*$ is negative at the saddle).]
    {\label{F4}\includegraphics[height=4.3cm,width=0.32\textwidth]{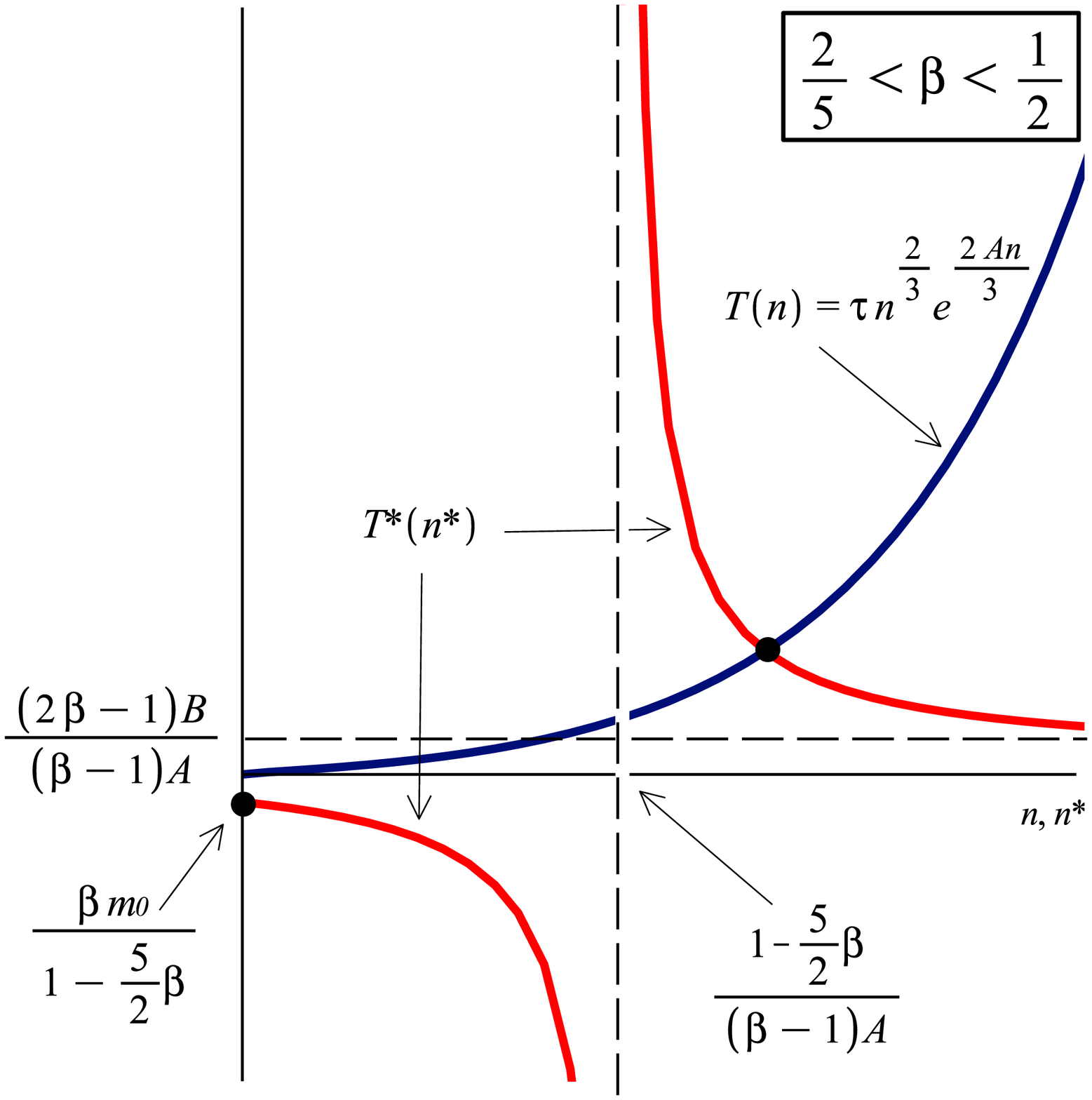}}

    \caption{\footnotesize{Parts (f) and (g) --- the case of $\frac{1}{2} < \beta < 1$. Also shown here --- in (h) --- is the separate case of $\frac{2}{5} < \beta < \frac{1}{2}$ which exhibits very similar behaviour to the case of $\frac{1}{2} < \beta < 1$, discussed on Figures 4b, 4c, 4d}}
    \label{Figure4_3}
    \end{figure}
    The only other curve that passes through the origin is the second integral $n = 0$. When $n_0 = 0$, the motion is restricted to the $H$-axis and is governed by $\dot{H} = -(3/2) H^2$ or $H(t) =  H_0 [1 + (3/2) H_0 (t - t_0)]^{-1}$.  If $H_0 > 0$, such trajectory converges to the origin along the $H$-axis in infinite time ($t \to \infty$). When $H_0 < 0$, such trajectories diverge to $H \to -\infty$ in time $t = t_0 + (2/3) \vert H_0 \vert^{-1}$. \\
    As no trajectory can cross $n = 0$ or the separatrix (the two curves, given by the second integrals), all trajectories between the separatrix and the $H$-axis in either of the half-planes [for each of these trajectories $\rho_d > 0$, namely, $I_2(n, H) = C > 0$] are therefore repelled by the origin. \\
    As discussed, the critical points of the two-component system (\ref{2d-1})--(\ref{2d-2}) have number densities which satisfy the equation $T^*(n^*) = T(n^*)$, namely:
    \b
    \label{starr}
    \frac{(2 \beta - 1) B n^* - \beta m_0}{(\beta - 1) A n^* + \frac{5}{2} \beta - 1} = \tau \, n^{*^{\frac{2}{3}}} \, e^{\frac{2 A n^*}{3}}.
    \e
    The analysis of $T^*(n^*)$ reveals several regimes --- see Figures 2 to 4 where all possibilities are shown. \\
    Extending the validity of the model for large values of $n$, one can investigate the blow-up $n \to \infty $ which occurs in finite time. Using the first integral $I_2(n, H, T) = C = $ const (\ref{i2_2}), for big $n$, the leading contribution in $H$ is
    \b
    H(n) & = & \pm \sqrt{ \frac{1}{2} \, \tau \,\, n^{\frac{5}{3}} \,\, e^{\frac{2 A n}{3}}},
    \e
    where $n(t)$ is again determined by separation of variables from $\dot{n} = -3 (1-\beta) H n$:
    \b
    \sqrt{\frac{2}{\tau}} \, \int n^{-\frac{11}{6}} \,\, e^{-\frac{ A n}{3}} \, dn = -3 \sigma (1 - \beta) (t - t^*).
    \e
    Here $\sigma = \mathrm{sign}(H)$ and $t^*$ is an integration constant. \\
    For $n \to \infty$, the integral behaves asymptotically as $-(3/A) \, n^{-11/6} \, e^{-An/3},$ thus:
    \b
    \label{bigcrunch}
    n^{-\frac{11}{6}}\,\,  e^{-\frac{An}{3}} = A \sigma (1 - \beta) \sqrt{\frac{\tau}{2}} \,\, (t - t^*).
    \e
    When $n \to \infty$, the left-hand side approaches zero and hence $t \to t^*$. Therefore, $t^*$ is the blow-up time and the above formulae are valid for $t<t^*$ only. This is clearly possible only when $(1 - \beta)\sigma < 0$ and sign$(H)$ = sign$(\beta - 1)$. Hence $H \to$ sign$(\beta-1)\infty $ and this blow-up represents a Big Crunch:
    \b
    n(t) \simeq -\frac{3}{A}\ln |t-t^*|.
    \e
One should also note that for $\beta > 1$, the regime of high $n$ and $H$ is characterized by inflation (see Figure 2c). The physical trajectories, when $\beta > 1$, are those with $H_0 > 0$ and they all diverge to $H \to \infty$ and $n \to \infty$ by getting very close to the separatrix as they do so. The leading term in $T(n)$ grows exponentially with $n$. Then $3H^2 \sim \rho \sim (3/2)nT$ and also $p \sim A n^2 T > 0$. Thus $\dot{H} = (\beta-1)(3H^2 +p)/2 > 0$ and $\ddot{a}/a= \dot{H} + H^2 >0$ which implies inflationary bahaviour.

\section{Conclusions}

The considered cosmological model has been reduced to a two-component autonomous nonlinear integrable dynamical system. This system however involves several physical parameters and, depending on these, its global behavior could be quite different, despite of the fact that the system is Hamiltonian and a conserved  Hamiltonian is identified. In physical terms this means that it describes various cosmological scenarios depending on the parameter choices. \\
The parameter choices and their implication for the global dynamics in terms of cosmological relevance are comprehensively studied and the physically meaningful parameter values are identified.  The presented examples illustrate all possible situations and in this sense a complete classification of the global behavior of the system is provided. \\
The (dynamically allowed) closed orbits and the saddles determine the essential behavior of the system, since these always appear in the spectrum of the Hamiltonian systems. In addition to the global conserved Hamiltonian, there are special (second) integrals, defined and conserved on a lower-dimensional manifold (lines or curves) in the two-dimensional phase space. They are invariant under the time evolution and separate the possible trajectories in the phase space. This further allows to identify specific sets of initial conditions in the phase space whose evolution is compliant with the fundamental laws (non-decreasing entropy, positive density and temperature). \\
The solution near the origin has been determined explicitly, as for example in (\ref{bigfreeze1}) and (\ref{bigfreeze2}), showing that the origin is reachable for an infinite time. The possibility for a blowup in finite time is also established in (\ref{bigcrunch}). \\
When $\beta > 1$, at high $n$ and $H$, the trajectories exhibit inflation --- driven by the process of matter creation. \\
The parameter $\beta$ is related to the rate of particle creation and is taken positive (by other authors as well). There is no fundamental principle that prevents the possibility of negative values of  $\beta$.  Indeed, for  $\beta$ and $H$ both negative the entropy increases and such situation is possible. It needs further investigation since the system is not invariant under the change of signs of both $\beta$ and $H$. \\
On the other hand, the system is symmetric under $n \to n$, $H \to - H$, and $t \to -t$, that is, the curves in the upper and the lower half-plane are symmetric, provided that the direction of the time arrows is reversed. In addition, one can study only trajectories with $n > 0$ since the line $n = 0$ is an invariant curve (second integral) and no trajectory can cross it, that is, trajectories starting at $n_0 < 0$ remain with $n < 0$ throughout their evolution, while those with $n_0 > 0$ remain with $n > 0$ throughout theirs.  All of the critical points are on the $H = 0$ axis, but the axis itself is not an invariant curve and the trajectories, in general, can cross from the upper half-plane into the lower half-plane or vice versa.
\begin{center}
\begin{sidewaystable}[ht]

\begin{tabular}{|c|c|c|c|c|c|c|c|}
\hline
\multirow{3}{*}{\begin{tabular}[c]{@{}c@{}}Equilibrium\\ Points\end{tabular}} & \multirow{3}{*}{Parameters} & \multicolumn{6}{c|}{$\beta$}                                                                                                                                                                                                                                                                                                                                                                                                                                                                                                                                              \\ \cline{3-8}
                                                                              &                             & \multicolumn{2}{c|}{$0 < \beta < \frac{2}{5}$}                                                                                                               & \multirow{2}{*}{$\frac{2}{5} < \beta < \frac{1}{2}$}                                        & \multicolumn{2}{c|}{$\frac{1}{2} < \beta < 1$}                                                                                                                                                                & \multirow{2}{*}{$\beta > 1$}                                                                 \\ \cline{3-4} \cline{6-7}
                                                                              &                             & $\frac{\beta m_0}{1-5\beta / 2} < \frac{(2 \beta - 1)B}{(\beta-1)A}$                  & $\frac{\beta m_0}{1-5\beta / 2} > \frac{(2 \beta - 1)B}{(\beta-1)A}$ &                                                                                             & $\frac{\beta m_0}{1-5\beta / 2} < \frac{(2 \beta - 1)B}{(\beta-1)A}$                                              & \multicolumn{1}{l|}{$\frac{\beta m_0}{1-5\beta / 2} > \frac{(2 \beta - 1)B}{(\beta-1)A}$} &                                                                                              \\ \hline
\multirow{2}{*}{$(n^*, H^*=0)$}                                               & $\tau > \tau_0$             & \begin{tabular}[c]{@{}c@{}}Two saddles\\ and a centre.\\ Fig. 3e, 3f, 3g\end{tabular} & \begin{tabular}[c]{@{}c@{}}One saddle.\\ Fig. 3a and 3c\end{tabular} & \begin{tabular}[c]{@{}c@{}}One saddle.\\ Fig. 4h.\\ See also\\ Fig. 4b, 4c, 4d\end{tabular} & \begin{tabular}[c]{@{}c@{}}Either one \\ saddle, or two \\ saddles and\\ a centre.\\ Fig. 4a, 4e, 4f\end{tabular} & \begin{tabular}[c]{@{}c@{}}One saddle.\\ Fig. 4a and 4c\end{tabular}                      & \begin{tabular}[c]{@{}c@{}}Does not exist.\\ Fig. 2b\end{tabular}                            \\ \cline{2-8}
                                                                              & $\tau < \tau_0$             & \begin{tabular}[c]{@{}c@{}}Two saddles\\ and a centre.\\ Fig. 3e, 3h, 3i\end{tabular} & \begin{tabular}[c]{@{}c@{}}One saddle.\\ Fig. 3a and 3d\end{tabular} & \begin{tabular}[c]{@{}c@{}}One saddle.\\ Fig. 4h.\\ See also\\ Fig. 4b, 4c, 4d\end{tabular} & \begin{tabular}[c]{@{}c@{}}Either one \\ saddle, or two \\ saddles and\\ a centre.\\ Fig. 4a, 4e, 4g\end{tabular} & \begin{tabular}[c]{@{}c@{}}One saddle.\\ Fig. 4a and 4d\end{tabular}                      & \begin{tabular}[c]{@{}c@{}}One saddle and \\ a centre.\\ Fig. 2c\end{tabular}                \\ \hline
\multirow{2}{*}{$(0,0)$}                                                      & $\tau > \tau_0$             & \multicolumn{6}{c|}{\begin{tabular}[c]{@{}c@{}}Attractive in the upper half-plane ($H > 0$),\\ repulsive in the lower half-plane ($H < 0$)\end{tabular}}                                                                                                                                                                                                                                                                                                                                                                                                                  \\ \cline{2-8}
                                                                              & $\tau < \tau_0$             & \multicolumn{5}{c|}{\begin{tabular}[c]{@{}c@{}}Attractive in the upper half-plane ($H > 0$),\\ repulsive in the lower half-plane ($H < 0$)\end{tabular}}                                                                                                                                                                                                                                                                                                                   & \begin{tabular}[c]{@{}c@{}}Attractive or repulsive\\ for different trajectories\end{tabular} \\ \hline
\end{tabular}
\end{sidewaystable}
\end{center}

    \end{document}